%% file: SUS-23-013_temp.tex
\documentclass[11pt,twoside,a4paper,cmspaper,final,collab]{cms-tdr}
\def\svnVersion{ed7084c-D}\def\svnDate{2026/04/22}\def\cmsCernNoTag{CERN-EP-2026-087}\def\cmsCernDate{\today}\def\cmsMessage{Submitted to the European Physical Journal C}
\begin{document}\cmsNoteHeader{SUS-23-013}

\newcommand{\Zjets}{\ensuremath{\PZ{+}\text{jets}}\xspace}
\newcommand{\Wjets}{\ensuremath{\PW{+}\text{jets}}\xspace}
\newcommand{\recoil}{\ensuremath{U}\xspace}
\newcommand{\recoilvec}{\ensuremath{\vec{\recoil}}\xspace}
\newcommand{\gq}{\ensuremath{g_\PQq}\xspace}
\newcommand{\gchi}{\ensuremath{g_\PZGc}\xspace}
\newcommand{\mchi}{\ensuremath{m_\PZGc}\xspace}
\newcommand{\darkHiggs}{\ensuremath{\PH_{\mathrm{D}}}\xspace}
\newcommand{\mzp}{\ensuremath{m_{\PZpr}}\xspace}
\newcommand{\Zvvjets}{\ensuremath{\PZ({\to}\PGn\PGn)}{+}\text{jets}\xspace}
\newcommand{\DYjets}{\ensuremath{\mathrm{DY}{+}\text{jets}}\xspace}
\newcommand{\Zlljets}{\ensuremath{\PZ({\to}\PGm\PGm,{\to}\Pe\Pe)}{+}\text{jets}\xspace}
\newcommand{\Wlvjets}{\ensuremath{\PW({\to}\PGm\PGn,{\to}\Pe\PGn)}{+}\text{jets}\xspace}

\newcommand{\mSD}{\ensuremath{m_{\mathrm{SD}}}\xspace}
\providecommand{\cmsTable}[1]{\resizebox{\textwidth}{!}{#1}}

\cmsNoteHeader{SUS-23-013}

\title{Search for dark matter produced in association with a dark Higgs boson decaying into a bottom quark-antiquark pair in proton-proton collisions at \texorpdfstring{$\sqrt{s}=13\TeV$}{sqrt(s) = 13 TeV}}

\date{\today}

\abstract{
    A search for dark matter produced in association with a dark Higgs boson decaying into a bottom quark-antiquark pair has been performed using proton-proton collision data at a center-of-mass energy of 13\TeV. The search uses data collected with the CMS detector at the CERN LHC during the 2016--2018 data-taking period, corresponding to an integrated luminosity of 138\fbinv. The results are interpreted in terms of a theoretical model of dark matter production that, together with a spin-1 gauge boson mediator, predicts the existence of a Higgs-boson-like particle in the dark sector (i.e., a dark Higgs boson). This search focuses on an experimental signature with large missing transverse momentum from dark matter production and a resonant structure in the invariant mass of the bottom quark-antiquark pair from the dark Higgs boson decay. Upper limits at 95\% confidence level on the signal strength for dark Higgs boson mass hypotheses below 160\GeV are set. Values of the mediator mass up to 4.5\,(2.5)\TeV are excluded at 95\% confidence level for a dark Higgs boson mass of 50\,(150)\GeV. This represents the most stringent limits set to date for the dark Higgs boson masses considered in this study.}

\hypersetup{
pdfauthor={CMS Collaboration},
pdftitle={Search for dark matter produced in association with a dark Higgs boson decaying into a bottom quark-antiquark pair in proton-proton collisions at sqrt(s)=13 TeV},
pdfsubject={CMS},
pdfkeywords={CMS, dark matter, Higgs boson, bottom quark}} 

\maketitle 

\section{Introduction}
\label{intro}

The predictions of the standard model (SM) of particle physics have been confirmed by decades of experiments. Despite these successes, the SM is still not able to explain phenomena such as the existence of dark matter (DM). While astrophysical observations have established that most of the matter in the universe is composed of DM~\cite{Arkani_Hamed_2009, 1937ApJ}, details of its nature remain elusive. 

One theoretically attractive model of DM is that of a thermally produced weakly interacting massive particle. The existence of such a particle with suitable mass and couplings could explain the abundance of DM in the universe, as well as many of the observed phenomena commonly ascribed to DM~\cite{Bertone:2004pz}. If nongravitational interactions exist between DM and SM particles, then the new interaction would imply the existence of a new mediator, and the Large Hadron Collider (LHC) at CERN would have the unique possibility of directly producing this mediating particle along with the DM particles.

Simplified models of DM production at the LHC~\cite{Abercrombie_2020} have become increasingly popular in recent years. These models assume that the pair production of DM particles in hadron collisions proceeds through a spin-0 or spin-1 bosonic mediator produced in the \textit{s}-channel. Such a mediator could be accompanied by a visible SM particle (or particles), often emitted as initial-state radiation (ISR). This gives rise to experimental signatures where the mediator decays into weakly interacting DM particles, appearing as an imbalance in the transverse momentum. Such signatures are commonly referred to as ``mono-X'', where X denotes either the SM particle produced in association with DM (such as a mono-photon, mono-W, or mono-Z) or its manifestation in the detector (such as a monojet). 

Among these signatures, monojet final states---in which an ISR gluon or quark is emitted and appears in the detector as a hadronic jet---offer a favorable topology due to the high rate of such an ISR emission. Monojet searches at the LHC~\cite{CMS:2021far,PhysRevD.103.112006} have strongly constrained the DM parameter space for models where DM relic particles in the cosmos annihilate directly into final-state SM particles. The resultant tension with astrophysical measurements of the abundance of DM is relaxed if the DM particles are not the lightest dark sector particles, leading to new annihilation channels. 

The minimal simplified model framework can be extended to include new DM annihilation channels via models in which, together with a spin-1 gauge boson \PZpr, a new complex Higgs field is introduced, whose vacuum expectation value spontaneously breaks the gauge symmetry in the dark sector~\cite{Duerr:2017uap}, giving rise to a new physical ``dark'' Higgs boson \darkHiggs. If the \darkHiggs boson is sufficiently light, DM particles can annihilate into a pair of dark Higgs bosons. This new annihilation channel would allow the model to easily match the observed relic abundance of DM~\cite{Bell:2016fqf,Kahlhoefer:2015bea,Bell:2016uhg}. In this model, the DM particle \PZGc is taken to be a Majorana fermion that couples axially to the gauge boson \PZpr. The \PZpr boson also has a vector-like coupling to SM quarks. The relevant part of the spin-1 sector of the model Lagrangian is:
\begin{equation*}
  \mathcal{L}_{\text{spin-1}} \;\supset\; -\,\gchi\,Z^\prime_\mu\,\overline{\chi}\,\gamma^\mu\gamma^5\,\chi \;-\; \gq\,Z^\prime_\mu\,\sum_{q}\,\bar{q}\,\gamma^\mu\,q,
\end{equation*}
where \gchi is the coupling between the \PZpr boson and the \PZGc particle, while \gq is the coupling between the \PZpr boson and the SM quarks. Following the LHC Dark Matter Working Group DM benchmark scenarios~\cite{Albert:2017onk}, these two parameters are set to 1.0 and 0.25, respectively.

\begin{figure}[!htbp]
    \centering
    \includegraphics[width=0.45\textwidth]{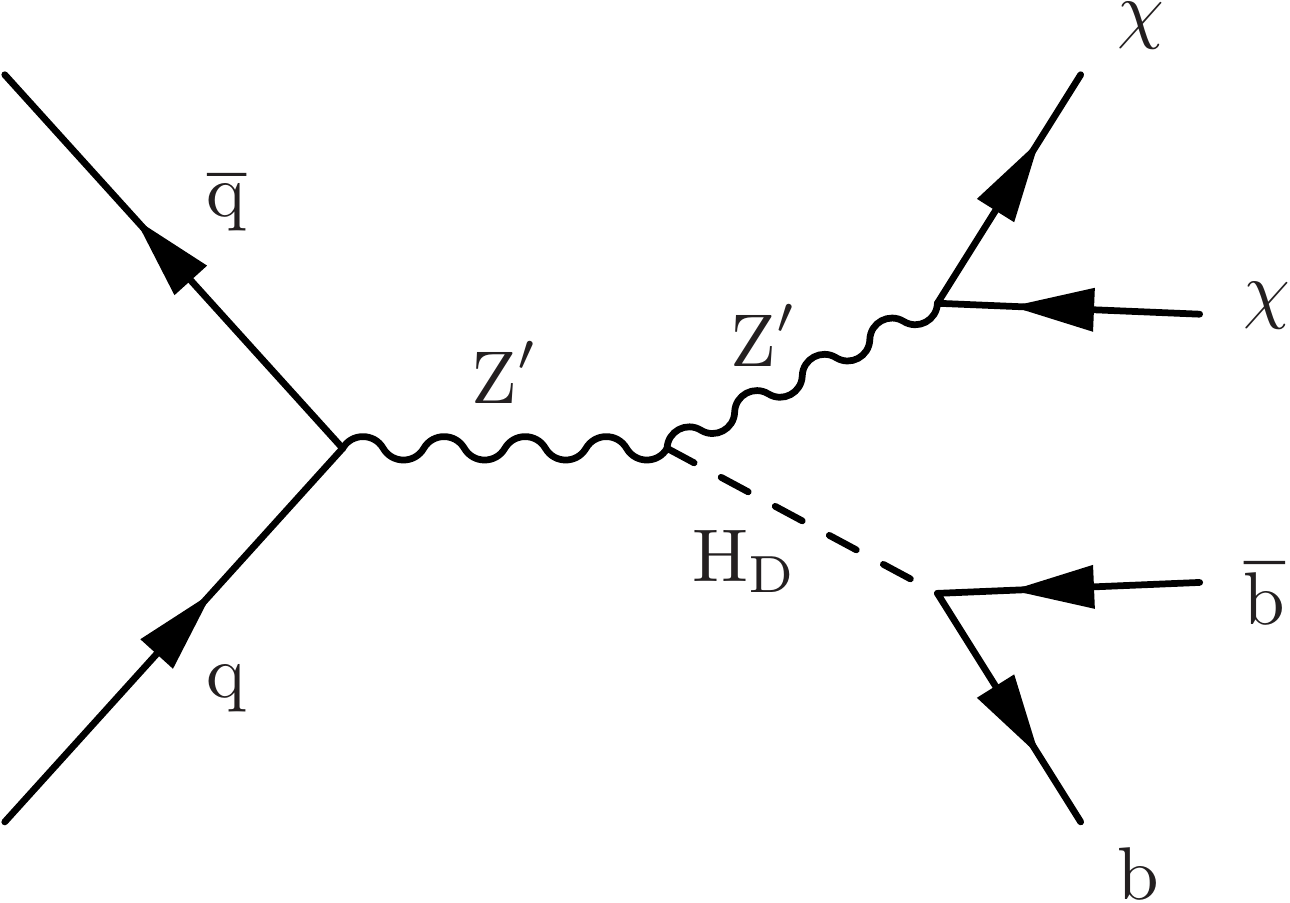}
    \caption{Feynman diagram for the associated production of an \darkHiggs boson and \PZGc particles.} 
    \label{fig:feyn}
\end{figure}

As the lightest state in the dark sector, the \darkHiggs boson does not decay into \PZGc particles, but it can decay into visible SM particles by mixing with the SM Higgs boson~\cite{Frandsen:2012rk,Kahlhoefer:2015bea}. For this reason, the decay into a bottom quark-antiquark pair (\bbbar) is expected to be dominant for \darkHiggs bosons with masses $m_{\darkHiggs} < 135\GeV$, and it is significant for $m_{\darkHiggs} < 160\GeV$. The mixing angle $\theta_{\mathrm{h}}$ between the \darkHiggs boson and the SM Higgs boson is set to 0.01, a value that is large enough to ensure prompt decay of the \darkHiggs boson while small enough to have no observable effect on the couplings of the SM Higgs boson~\cite{CMS:2022dwd}. This motivates searches for DM in collisions where \PZGc particles are produced in association with an \darkHiggs boson decaying into a \bbbar pair. The production mechanism, illustrated in Fig.~\ref{fig:feyn}, proceeds through quark-antiquark annihilation into a \PZpr boson, which emits an \darkHiggs boson via dark Higgsstrahlung before decaying into a pair of \PZGc particles.

In this paper, we present a search for such a production mechanism in events with large missing transverse momentum (\ptmiss) from DM particles, accompanied by a large-cone jet consistent with the hadronization of a \bbbar pair from the decay of an \darkHiggs boson. Searches for \darkHiggs bosons produced in association with DM have already been performed by the ATLAS~\cite{ATLAS:2022bzt} and CMS~\cite{CMS:2023dof} Collaborations with data collected during the 2016--2018 data-taking period. These searches focus on heavier $m_{\darkHiggs}$ hypotheses, larger than 160\GeV. For an \darkHiggs boson of such a mass, the decay into a pair of \PW bosons is dominant. The ATLAS Collaboration has also recently published a search for lower-mass dark Higgs bosons decaying into a \bbbar pair~\cite{ATLAS:2024ypx}. In this paper, we describe a similar search that uses the data set collected by the CMS experiment at a center-of-mass energy of 13\TeV during the 2016--2018 data-taking period, corresponding to an integrated luminosity of 138\fbinv.  Tabulated results are provided in the HEPData record for this analysis~\cite{hepdata}.

\section{The CMS detector and event reconstruction}
\label{objects}

The CMS apparatus~\cite{Chatrchyan:2008zzk,CMS:2023gfb} is a multipurpose, nearly hermetic detector, designed to trigger on~\cite{Chatrchyan:2008zzk,CMS:2023gfb} and identify electrons, muons, photons, and (charged and neutral) hadrons~\cite{Chatrchyan:2008zzk,CMS:2023gfb}. Its central feature is a superconducting solenoid of 6\unit{m} internal diameter, providing a magnetic field of 3.8\unit{T}. Inside the solenoid volume are a silicon pixel and strip tracker, a lead tungstate crystal electromagnetic calorimeter (ECAL), and a brass and scintillator hadron calorimeter (HCAL), each composed of a barrel and two endcap sections. Forward calorimeters extend the pseudorapidity $(\eta)$ coverage provided by the barrel and endcap detectors. Muons are reconstructed using gas-ionization detectors embedded in the steel flux-return yoke outside the solenoid. More detailed descriptions of the CMS detector, together with a definition of the coordinate system used and the relevant kinematic variables, can be found in Refs.~\cite{Chatrchyan:2008zzk,CMS:2023gfb}.

The silicon tracker used in 2016 measured charged particles in the range $\abs{\eta} < 2.5$. For non-isolated particles of $1 < \pt < 10\GeV$ and $\abs{\eta} < 1.4$, the track resolutions were typically 1.5\% in \pt and 25--90 (45--150)\mum in the transverse (longitudinal) impact parameter~\cite{TRK-11-001}. At the beginning of 2017, a new pixel detector was installed~\cite{Phase1Pixel}; the upgraded tracker measured particles up to $\abs{\eta} = 3.0$ with typical resolutions of 1.5\% in \pt and 20--75\mum in the transverse impact parameter~\cite{DP-2020-049} for non-isolated particles of $1 < \pt < 10\GeV$. The primary vertex is taken to be the vertex corresponding to the hardest scattering in the event, evaluated using tracking information alone, as described in Section 9.4.1 of Ref.~\cite{CMS-TDR-15-02}. 

In the region $\abs{\eta} < 1.74$, the HCAL cells have widths of 0.087 in $\eta$ and 0.087 radians in azimuth ($\phi$). In the $\eta$-$\phi$ plane, and for $\abs{\eta} < 1.48$, HCAL cells map to $5{\times}5$ arrays of ECAL crystals to form calorimeter towers projecting radially outward from close to the nominal interaction point. For $\abs{\eta} > 1.74$, the coverage of the towers increases progressively to a maximum of 0.174 in $\Delta \eta$ and $\Delta \phi$. The forward hadron (HF) calorimeter uses steel as an absorber and quartz fibers as the sensitive material. The two halves of the HF are located 11.2\unit{m} from the interaction region, one at each end, and together they provide coverage in the range $3.0 < \abs{\eta} < 5.2$. They also serve as luminosity monitors.  

Events of interest are selected using a two-tier trigger system. The first level (L1), composed of custom hardware processors, uses information collected from the calorimeters and muon detectors to select events at a rate of around 100\unit{kHz} within a fixed latency of 4\mus~\cite{Sirunyan:2020zal}. The second level, known as the high-level trigger (HLT), consists of a farm of processors running a version of the full event reconstruction software optimized for fast processing, and reduces the event rate to a few kHz before data storage~\cite{Khachatryan:2016bia,CMS:2024aqx}.

A particle-flow (PF) algorithm~\cite{CMS:2017yfk} aims to reconstruct and identify each individual particle in an event, with an optimized combination of information from the various elements of the CMS detector. In this process, the identification of the PF candidate type (photon, electron, muon, charged and neutral hadrons) plays an important role in the determination of the particle direction and energy. The energy of the photons is obtained from the ECAL measurement. The energy of electrons is determined from a combination of the electron momentum at the primary interaction vertex as determined by the tracker, the energy of the corresponding ECAL cluster, and the energy sum of all of the bremsstrahlung photons spatially compatible with originating from the electron track. The energy of muons is obtained from the curvature of the corresponding track. The energy of charged hadrons is determined from a combination of their momentum measured in the tracker and the matching ECAL and HCAL energy deposits, corrected for the response function of the calorimeters to hadronic showers. The energy of neutral hadrons is obtained from the corresponding corrected ECAL and HCAL energies. 

In this search, electrons (photons) are required to have $\pt>10$ (15)\GeV and $\abs{\eta}<2.5$. Muons are required to have $\pt>10\GeV$ and $\abs{\eta}<2.4$. All leptons and photons are required to be isolated. Isolation is calculated by imposing thresholds on the energy of PF candidates within a certain distance $\Delta R = \sqrt{\smash[b]{(\Delta\eta)^{2} + (\Delta\phi)^{2}}}$ with respect to the lepton or photon. The PF candidates with $\Delta R < 0.3$ are considered when computing the isolation of electrons and photons, while $\Delta R < 0.4$ is used for muons.
Additional selection criteria are applied to define ``veto'' electrons, as well as ``loose'' muons and photons~\cite{CMS:2020uim,Sirunyan:2018fpa}, which are used to reject unwanted events. 
Electrons and muons used to select events in control data samples are instead required to satisfy ``tight'' selection criteria. The threshold requirement on the $\pt$ is also increased to 40 (20)\GeV for ``tight'' electrons (muons).

Hadronically decaying tau leptons are first required to pass identification criteria using the hadron-plus-strips algorithm~\cite{Khachatryan:2015dfa}. Subsequently, a deep neural network, called \textsc{DeepTau}~\cite{deeptau}, is used to discriminate genuine hadronic tau leptons against misidentified jets, electrons, and muons. In addition, the overlap between tau leptons and electrons or muons is further suppressed by removing tau leptons that lie within $\Delta R < 0.4$ of a well-reconstructed and isolated electron or muon. Hadronically decaying tau leptons are required to have $\pt>20\GeV$ and $\abs{\eta}<2.3$.

For each event, hadronic jets are clustered from the PF candidates using the infrared and collinear safe anti-\kt algorithm~\cite{Cacciari:2008gp, Cacciari:fastjet1} with a distance parameter of 0.4 or 1.5. 
Depending on the respective distance parameter, these jets are referred to as ``AK4'' or ``AK15'' jets.
Jet momentum is determined as the vectorial sum of all particle momenta in the jet, and is found from simulation to be, on average, within 5--10\% of the true momentum over the entire \pt spectrum and detector acceptance. Jet energy corrections are derived from simulation to bring the measured response of jets to that of particle-level jets on average. In situ measurements of momentum balance in dijet, $\text{photon} + \text{jet}$, $\PZ + \text{jet}$, and multijet events are used to account for any residual differences in the jet energy scale between data and simulation~\cite{Khachatryan:2016kdb}. The jet energy resolution typically amounts to 15--20\% at 30\GeV, 10\% at 100\GeV, and 5\% at 1\TeV~\cite{Khachatryan:2016kdb}. Additional selection criteria are applied to each jet to remove jets potentially dominated by anomalous contributions from various subdetector components or reconstruction failures~\cite{CMS:2017wyc}.

Additional proton-proton collisions within the same or nearby bunch crossings (pileup) can contribute additional tracks and calorimetric energy depositions to the jet momentum.
To mitigate this effect in the AK4 jets, charged particles identified to be originating from pileup vertices are discarded, and an offset correction is applied to correct for remaining contributions~\cite{Khachatryan:2016kdb}. For AK15 jets, the pileup-per-particle identification algorithm~\cite{Sirunyan:2020foa,puppi} is used to mitigate the effect of pileup at the reconstructed particle level, making use of local shape information, event pileup properties, and tracking information. Specifically, a local shape variable that distinguishes between collinear and soft diffuse distributions of other particles surrounding the particle under consideration is defined. The former is attributed to particles originating from the hard scatter, and the latter is attributed to particles originating from pileup interactions. Charged particles identified to be originating from pileup vertices are discarded. For each neutral particle, the local shape variable is computed using the surrounding charged particles compatible with the primary vertex within the tracker acceptance ($\abs{\eta} < 2.5$), and using both charged and neutral particles in the region outside of the tracker coverage. The momenta of the neutral particles are then rescaled according to their probability to originate from the primary interaction vertex deduced from the local shape variable, superseding the need for jet-based pileup corrections~\cite{Sirunyan:2020foa}.

The AK4 jets used in this search are further required to have a $\pt > 30\GeV$ and $\abs{\eta} < 2.5$. Jets with $\Delta R < 0.4$ with respect to a well-identified and isolated lepton or photon are removed. To identify jets originated by the hadronization of \PQb quarks (hereafter referred to as ``\PQb jets''), the \textsc{DeepJet} algorithm~\cite{Bols:2020bkb} is employed. A loose working point is used, defined for each data-taking year as the minimum requirement in the \textsc{DeepJet} discriminator distribution, which returns a 10\% rate of misidentifying a jet originated by a light-flavor quark or gluon. The loose working point corresponds to an efficiency of correctly identifying jets originated by \PQb quarks (i.e., \PQb tagging efficiency) of 90--95\%, depending on the \pt of the AK4 jet~\cite{CMS-DP-2023-005}.

The AK15 jets used in this search are further required to have a \pt larger than 160\GeV and $\abs{\eta} < 2.4$. Jets with $\Delta R < 1.5$ with respect to a well-identified and isolated lepton or photon are removed. The modified mass drop tagger algorithm~\cite{Dasgupta:2013ihk,Butterworth:2008iy}, also known as the soft-drop (SD) algorithm~\cite{msd}, with the angular exponent $\beta = 0$, soft cutoff threshold $z_{\text{cut}} < 0.1$, and radius $R_{0} = 1.5$, is applied to remove soft, wide-angle radiation from the jet. 
To identify AK15 jets that are consistent with the hadronization of a $\PQb\PAQb$ pair from the decay of a Lorentz-boosted massive resonance, a deep neural network \textsc{DeepAK15} is used~\cite{Sirunyan:2020lcu}. More specifically, the ``mass-decorrelated'' version of \textsc{DeepAK15} is employed. In this variant, we use adversarial training: a second neural network (`adversary') is trained to predict the AK15 jet mass from the \textsc{DeepAK15} outputs. The joint loss includes a penalty term that grows with the adversary's performance. Therefore, this method optimizes the ability to correctly identify the origin of an AK15 jet while systematically decorrelating the output score from the AK15 jet mass. This approach avoids shaping the AK15 jet mass distribution in background events. Since the strategy for this search relies on the identification of a peak in the AK15 jet mass distribution due to a resonance of unknown mass, the mass-decorrelated version of the tagger offers the best option.

The vector \ptvecmiss is computed as the negative vector \pt sum of all the PF candidates in an event, and its magnitude is denoted as \ptmiss. The \ptvecmiss is modified to account for corrections to the energy scale and resolution of the reconstructed AK4 jets in the event~\cite{Sirunyan:2019kia}.  Anomalous high-\ptmiss events can be due to a variety of reconstruction failures, detector malfunctions, or non-collision backgrounds. Such events are rejected by dedicated filters that are designed to eliminate more than 85--90\% of spurious high-\ptmiss events with a signal efficiency close to 100\%~\cite{Sirunyan:2019kia}. 

The vector \recoilvec is defined as the vector sum of \ptvecmiss and the \ptvec of any identified leptons in the event. Its magnitude is denoted as the hadronic recoil \recoil. It represents the total \pt of all the nonhadronic particles in each event. The signal events in this search contain only jets and no other reconstructed candidates, therefore \recoil is equivalent to \ptmiss of the event. Because of the presence of DM particles, large \recoil is expected and therefore its distribution is used, together with the AK15 jet mass, to distinguish signal from the background events. For the leading background processes, identified in Section~\ref{selection}, \recoil corresponds to the \pt of a vector boson. In those data samples of this search where no lepton is required, the lepton from the vector boson decay is missed and \recoil is once again equivalent to the \ptmiss of the event. In data samples where a lepton is required in the final state, the lepton from the vector boson decay is identified and \recoil is equivalent to the magnitude of the vector sum of \ptvecmiss and the lepton \ptvec.

\section{Simulated samples}

Samples of Monte Carlo simulated events are used to predict the signal and background contributions. In all cases, parton showering, hadronization, and underlying event properties are modeled using \PYTHIA~\cite{Sjostrand:2014zea} version 8.202 or later with the underlying event tune CUETP8M1 or CP5~\cite{CMS:2019csb}, based on the data-taking year. 
The simulation of the interactions between the particles and the CMS detector is based on \GEANTfour~\cite{Agostinelli:2002hh}. The NNPDF 3.0~\cite{Ball2015} and NNPDF 3.1~\cite{Ball:2017nwa} next-to-next-to-leading order (NNLO) parton distribution functions are used for the generation of all samples based on the year of data collection. The same reconstruction algorithms used for the data are applied to the simulated samples. 

This analysis considers the following SM vector boson plus jets processes: \Zvvjets, referred to as \Zjets; \Zlljets, referred to as \DYjets; and \Wlvjets, referred to as \Wjets. These processes are simulated with up to two additional partons at leading order (LO) in quantum chromodynamics (QCD) using \MGvATNLO \cite{MADGRAPH} v2.4.2 and normalized to their next-to-LO (NLO) cross section. The MLM scheme~\cite{Mangano:2006rw} is used to match matrix element partons to the parton shower. Samples of events with top quark pairs (\ttbar production) are generated at NLO in QCD, with up to two additional partons in the matrix element calculation, using \MGvATNLO version 2.6.5 in conjunction with the FxFx jet matching scheme~\cite{Frederix:2012ps}. Their cross sections are normalized to the inclusive cross section of \ttbar production at NNLO in QCD~\cite{Czakon:2013goa}. 
Events with electroweakly produced single top quarks (single \PQt production) are simulated using \POWHEG 2.0~\cite{Nason:2004rx,Frixione:2007vw,Alioli:2010xd,Alioli:2009je,Re:2010bp} and normalized to the inclusive cross section calculated at NLO in QCD~\cite{Aliev:2010zk,Kant:2014oha}.

The associated production of vector bosons ($\PV\PV$ production) is simulated at LO using \PYTHIA, and normalized to the cross sections at NNLO precision for $\PW\PW$ production~\cite{Gehrmann:2014fva}, and at NLO precision for $\PW\PZ$ and $\PZ\PZ$ production~\cite{Campbell:1999ah}. Several production mechanisms of SM Higgs bosons decaying into \PQb quark pairs ($\PH {\to} \bbbar$ production) are also produced at LO with the \POWHEG generator. Samples of QCD multijet production events are generated at LO using \MGvATNLO.

Simulated samples of \darkHiggs boson production are generated with \MGvATNLO~\cite{MADGRAPH} at LO, including up to one additional parton in the matrix element calculations, with the MLM matching scheme. Separate samples are generated for different mass hypotheses for the \PZpr boson, \PZGc particles, and the \darkHiggs boson.

\section{Event selection}
\label{selection}

The signal region (SR) in this analysis is characterized by a large \recoil due to DM production and the presence of an AK15 jet consistent with an $\darkHiggs {\to} \bbbar$ decay.

The SR trigger selections require $\ptmiss > 120\GeV$ and $\mht > 120\GeV$, where \ptmiss is calculated using all PF candidates reconstructed at the HLT, except for the muons, and where \mht is defined as the magnitude of the negative vector \ptvec sum of all AK4 hadronic jets in the event reconstructed at the HLT.  

After the trigger selection, events in the SR are required to satisfy $\recoil > 250\GeV$. The soft-drop corrected mass ($\mSD$) of the leading AK15 jet in \pt must satisfy the requirement $\mSD\in[40,300]$. To enhance the selection of events where the leading AK15 jet arises from the hadronic decay of an \darkHiggs boson the jet is required to pass a minimum requirement in the  \textsc{DeepAK15} score. The threshold varies by data-taking year to optimize signal sensitivity, corresponding to an average signal efficiency of 90--95\%

After this basic pre-selection, the main background processes in this search are from \Zjets, \Wjets, \ttbar, and QCD multijet production. The \Zjets process is the largest background and is irreducible. In contrast, the background from \Wjets and \ttbar processes is suppressed by rejecting events with a well-reconstructed and isolated lepton. The ``veto'' working point is used in the identification of electrons, while the ``loose'' working point is used in the identification of muons. Similarly, events with hadronically decaying tau leptons are rejected. To help suppress electroweak (EW) backgrounds with an ISR photon, events containing a ``loose'' photon are also rejected. The \ttbar process is further suppressed by vetoing events with \PQb-tagged AK4 jets that do not overlap with the leading AK15 jet. Non-overlapping AK4 jets are identified by imposing a requirement on the $\Delta R$ with the leading AK15 jet to be larger than 1.5. 
To reject QCD multijet events with large \recoil arising from mismeasurements of the jet momenta, it requires the $\Delta\phi >$ 0.5 (1.5) radians between the \recoilvec direction and each AK4 (AK15) jet in the event.

To avoid events with anomalous \recoil due to reconstruction failures of the PF algorithm, events are required to have $\abs{\ptmiss(\mathrm{PF}) - \ptmiss(\text{calorimeter})}/\recoil < 0.5$, where $\ptmiss(\mathrm{PF})$ refers to the standard \ptmiss computed from PF candidates as defined in Section~\ref{objects}, while $\ptmiss(\text{calorimeter})$ is the \ptmiss calculated using only information from the calorimeters. To mitigate mismeasurement from a nonfunctioning HCAL section in the 2018 data~\cite{CMS:2023gfb}, events with jets in that region or with $\phi(\ptvecmiss)\in[-1.62,-0.62]$ at $\ptmiss < 470\GeV$ are rejected.

As shown in Table~\ref{sr_yields}, even after the full SR selection is applied, the selected data sample still has a large contamination from \Zjets, \Wjets, and \ttbar production. In order to characterize and constrain these background processes, dedicated control regions (CRs) are used. 

\setlength{\tabcolsep}{3pt} 
\begin{table*}[!htbp]
    \centering
    \def\arraystretch{1}
    \topcaption{Expected yields from background processes in the SR\@. The values shown are from Monte Carlo simulation, and the uncertainties are statistical only. The expected yields for a reference signal hypothesis with $\mzp = 1000\GeV$, $m_{\darkHiggs} = 130\GeV$, and $\mchi = 150\GeV$ are also reported.}
    \label{sr_yields}
    \newcolumntype{x}{D{,}{\,\pm\,}{5.4}}
    \begin{tabular}{lxxx}
        \hline
         & \multicolumn{1}{c}{2016} & \multicolumn{1}{c}{2017} & \multicolumn{1}{c}{2018} \\
         \hline
         \Zjets       & 7515,29 & 7035,33 & 6979,39 \\
         \ttbar                         & 5487,200 & 5811,60 & 6784,130 \\
         \Wjets      & 3998,39 & 2991,40 & 2827,50 \\
         \PV\PV                            & 718,18 & 623,18   & 606,21 \\
         Single \PQt                           & 646,11 & 567,13   & 615,13 \\
         QCD multijet                       & 93,26 & 155,42   & 163,65 \\
         $\PH {\to} \bbbar$            & 57.58,0.34 & 72.03,0.33       & 83.79,0.34 \\
         \DYjets     & 56.8,2.2 & 43.3,2.0       & 37.1,3.0 \\
      \hline
    Total background                     & 18571,210 & 17298,92 & 18094,160 \\[3pt]
    $\mzp = 1000\GeV$, $m_{\darkHiggs} = 130\GeV$, $\mchi = 150\GeV$              & 2329,14 & 2132,13 & 2337,16 \\[1pt]
      \hline
    \end{tabular}
\end{table*}

A CR enriched in \Zjets events (ZCR) is identified using the same requirements that define the SR, but inverting the criterion on the \textsc{DeepAK15} score.

Similarly, CRs enriched in \Wjets events are identified using the full set of SR selection criteria with the exception of the muon or electron veto. More specifically, single-muon CRs are composed of events with exactly one tight muon, and single-electron CRs are composed of events with exactly one tight electron. Since the events in both CRs ``pass'' the requirement in the minimum \textsc{DeepAK15} score, the CRs are labeled as WMPCR and WEPCR respectively. An additional set of single-lepton CRs is used, composed of events that satisfy the same requirements that define the WMPCR and WEPCR, but ``fail'' to meet the \textsc{DeepAK15} score criterion on the leading AK15 jet. They are therefore labeled as WMFCR and WEFCR\@. At the HLT, muons are excluded from the \ptmiss calculation and therefore events with high-\pt muons show large \ptmiss. For this reason the same triggers based on \ptmiss and \mht used for the SR are also used to select events that populate the single-muon CRs. The control samples with electrons are selected using two single-electron triggers: one that requires $\pt>27$ (2016), 35 (2017), 32 (2018)\GeV, while the other requires $\pt>105$ (2016) and 115\GeV (2017--2018). Single-electron triggers differ in their isolation requirements: while the lower-threshold trigger requires electrons to be well isolated, the higher threshold trigger does not, which improves the efficiency at high \pt. Additionally, a single-photon trigger with $\pt>200\GeV$ is used in 2017--2018, which avoids reliance on the HLT track reconstruction and increases the overall efficiency for electrons with $\pt > 200\GeV$. During the 2016--2017 data-taking years, a gradual shift in the timing of the inputs of the ECAL L1 trigger in the region $\abs{\eta} > 2.0$ caused a specific trigger inefficiency, known as ``L1 prefiring''. Correction factors are computed from the data and applied to the acceptance evaluated by simulation for the 2016--2017 samples. Although the same requirement on \recoil that defines the SR is used in all single-lepton CRs, the \ptmiss in each event of the single-electron regions is required to be larger than 100\GeV, in order to suppress the larger contribution from QCD multijet events due to jets misidentified as electrons. 

Dedicated single-lepton CRs are also used to constrain the background from \ttbar production. The same single-electron (TTECR) or single-muon (TTMCR) selections used to define the WMPCR and WEPCR are applied, but the veto on b-tagged AK4 jets that do not overlap with the leading AK15 jet is inverted. In this case, only CRs populated with events that satisfy the leading AK15 jet \textsc{DeepAK15} score requirement are used. Table~\ref{tab:selection} summarizes the requirements that define the SR and CRs.

\setlength{\tabcolsep}{4pt} 
\begin{table*}[!htbp]
    \centering
    \def\arraystretch{1}
    \topcaption{Summary of requirements that define the different analysis regions. A ``\checkmark'' means the requirement is enforced; a ``$\times$'' means the variable is left unconstrained in that region (the requirement is not applied).}
   \label{tab:selection}
  \cmsTable{
   \begin{tabular}{l c c c c c c c c}
        \hline
         Selection & SR & ZCR & WMPCR & WEPCR & WMFCR & WEFCR & TTMCR & TTECR  \\
          \hline
      \ptmiss+\mht trigger & \checkmark & \checkmark & \checkmark & $\times$ & \checkmark & $\times$ & \checkmark & $\times$ \\
      Single-electron trigger & $\times$ & $\times$ & $\times$ & \checkmark & $\times$ & \checkmark & $\times$ & \checkmark \\
      $\recoil > 250\GeV$ & \checkmark & \checkmark & \checkmark & \checkmark & \checkmark & \checkmark & \checkmark & \checkmark \\
      $\abs{\ptmiss(\mathrm{PF}) - \ptmiss(\text{calo})}/\recoil <$ 0.5 & \checkmark & \checkmark & \checkmark & \checkmark & \checkmark & \checkmark & \checkmark & \checkmark \\
       AK15 $\pt > 160\GeV$ & \checkmark & \checkmark & \checkmark & \checkmark & \checkmark & \checkmark & \checkmark & \checkmark \\
    AK15 \mSD $\in$ [40, 300]\GeV & \checkmark & \checkmark & \checkmark & \checkmark & \checkmark & \checkmark & \checkmark & \checkmark \\
    min $\Delta\phi(\vec{U},\vec{\mathrm{AK4s}}) > 0.5$  & \checkmark & \checkmark & \checkmark & \checkmark & \checkmark & \checkmark & \checkmark & \checkmark \\
    min $\Delta\phi(\vec{U},\vec{\mathrm{AK15s}}) > 1.5$  & \checkmark & \checkmark & \checkmark & \checkmark & \checkmark & \checkmark & \checkmark & \checkmark \\
    $\ptmiss > 100\GeV$  & $\times$ & $\times$ & $\times$ & \checkmark & $\times$ & \checkmark & $\times$ & \checkmark \\
    \# of muons & 0 & 0 & 1 & 0 & 1 & 0 & 1 & 0 \\
    \# of electrons & 0 & 0 & 0 & 1 & 0 & 1 & 0 & 1 \\
    \# of photons & 0 & 0 & 0 & 0 & 0 & 0 & 0 & 0 \\
    \# of taus & 0 & 0 & 0 & 0 & 0 & 0 & 0 & 0 \\
    \# of extra b-tagged AK4 jets & 0 & 0 & 0 & 0 & 0 & 0 & $\geq$1 & $\geq$1 \\
    \textsc{DeepAK15} requirement & pass & fail & pass & pass & fail & fail & pass & pass \\
      \hline
   \end{tabular}
   }
\end{table*}

\section{Background estimation}

Background estimation and signal extraction are performed simultaneously, using a joint maximum likelihood fit across the SR and all of the CRs for each data-taking year. A likelihood function is constructed to model the expected background contributions in each bin of the two-dimensional \recoil-vs.-$\mSD$ variable of the SR and CRs, as well as the expected signal yield in each bin of the SR\@. The best fit background model, as well as the best fit signal strength modifier $\mu$ (which---for a given signal hypothesis---controls the signal normalization relative to the theoretical cross section), are obtained by maximizing this likelihood function.

Predictions for the dominant backgrounds \Zjets, \Wjets, and \ttbar in the SR are based on the yield of the same processes in each bin of the CRs. The per-bin yields for these processes in the SR are defined as free parameters of the likelihood function. A different set of free parameters is used for each data-taking year. The yields in the CRs are then defined relative to these parameters by introducing a set of per-bin transfer factors. The choice of transfer factors takes into account the correlations among the \Zjets and \Wjets background contributions in all regions. In all cases, the central values of the transfer factors are obtained from the ratios of the simulated \recoil-vs.-$\mSD$ spectra of the respective processes in the SR to those in the CRs. The predictions for the subdominant single t, $\PV\PV$, $\PH {\to} \bbbar$, \DYjets, and QCD multijet backgrounds in the SR and CRs are taken directly from simulation.

To achieve the most accurate possible predictions of the ratios between the dominant backgrounds in the SR and CRs, as well as of the absolute normalization and shape of the \recoil-vs.-$\mSD$ distributions for the subdominant backgrounds, corrections are applied to each simulated event. These corrections take into account both experimental and theoretical effects that are not present in the simulated samples. The experimental corrections are related to the trigger efficiencies, the identification and reconstruction efficiencies of charged leptons, the efficiencies of the \textsc{DeepJet} and \textsc{DeepAK15} algorithms, and the pileup distribution in simulation. Theoretical corrections are applied to the \Zjets and \Wjets processes to model the effects of NLO terms in the perturbative EW corrections~\cite{DMTheory}. The corrections are parameterized as functions of the generator-level boson \pt and are evaluated separately for the \Wjets and \Zjets processes.

\section{Systematic uncertainties}

Systematic uncertainties are incorporated into the likelihood function as nuisance parameters. In the case of the \Zjets, \Wjets, and \ttbar processes, the nuisance parameters affect the values of the transfer factors in each bin of the \recoil-vs.-$\mSD$ variable and thus control the ratios of the contributions from different processes, as well as the ratios of the yields in the SRs to those in various CRs. For the subdominant background processes, the yields in each bin are directly parameterized in terms of the nuisance parameters. 

Uncertainties in the measurement of the integrated luminosity in each data-taking year are 0.6--2.0\%~\cite{CMS:2021xjt,CMS-PAS-LUM-17-004,CMS-PAS-LUM-18-002}, with an overall uncertainty for the 2016--2018 data of 1.6\%. The uncertainties in the corrections for the L1 pre-firing effect in 2016--2017 as well as the uncertainties in the pileup correction are of the order of 1\%.
The uncertainties in the efficiencies of reconstructing and identifying electron candidates are 1\% and 2--3\%, respectively. For muons, the uncertainties in the identification efficiency are 1\%, with an additional 1\% uncertainty in the efficiency of the isolation criteria.
A systematic uncertainty for each lepton or photon veto selection has been obtained by propagating the overall uncertainties in the identification of muons, electrons, photons, and tau leptons into the vetoed regions. While the uncertainties are found to be negligible for the photon, muon, and electron vetoes, a 3\% uncertainty in the tau lepton veto is included.
The uncertainties in the trigger efficiency are 1\% for the single-electron trigger and 1--2\% for the \ptmiss trigger.
The uncertainty in the modeling of \ptmiss in simulation~\cite{Khachatryan:2014gga} is dominated by the uncertainty in the jet energy corrections. 

Application of the above uncertainties leads to bin migrations, which in turn lead to a variation in the rate of events passing the minimum requirement on \recoil. The resulting change in rate is estimated to be 5\%, and is included as a systematic uncertainty. The corresponding bin migration effects in the jet \pt spectrum that result from the uncertainties in the AK15 jet energy corrections lead to an estimated 4\%  variation in the rate of events passing the minimum AK15 jet \pt requirement. The uncertainty in the \textsc{DeepJet} efficiency leads to a shape uncertainty, which is applied to all processes in all regions. The uncertainty in the \textsc{DeepAK15} efficiency results in a shape uncertainty, which is applied to the signal processes in the SR\@. 

Uncertainties of 100\% are assigned to the normalization of the QCD multijet background contributions in all regions. These uncertainties are correlated among the regions with the same source of misidentified leptons. In order to account for these correlations, an uncertainty is applied to QCD multijet events in the SR and CR enriched in \Zjets events, while a separate uncertainty is applied to QCD multijet events in the single-muon CRs, and another uncertainty is applied to the single-electron CRs. Additionally, uncertainties of 20\% are assigned to the cross sections of $\PV\PV$, $\PH {\to} \bbbar$, and \DYjets production. Similarly, 10\% uncertainties in the single \PQt and \ttbar production cross sections are also assigned. 

The theoretical uncertainties in the transfer factors related to higher-order effects in the QCD and EW perturbative expansions are calculated according to the prescription given in Ref.~\cite{DMTheory}, and implemented as described in Ref.~\cite{Sirunyan:2017jix}. 
Bin-by-bin statistical uncertainties are incorporated following the Barlow--Beeston ``lite'' approach~\cite{Conway:2011in}.

The likelihood functions obtained for the three data-taking years are combined. The combination is performed by defining a combined likelihood that describes all the regions in all data sets. For this purpose, the effects of all theoretical uncertainties are assumed to be correlated. Most experimental uncertainties are dominated by the inherent precision of auxiliary measurements specific to each data set and are thus assumed to be uncorrelated among the different data-taking years. The experimental uncertainties related to the determination of the integrated luminosity and to the \textsc{DeepJet} efficiency are partially correlated among the data-taking years, which is taken into account by splitting the total uncertainty into its correlated and uncorrelated components. A summary of all the uncertainties considered for this analysis is presented in Table~\ref{tab:systematics}.

\begin{table}[!htbp]
\centering
    \def\arraystretch{1.2}
    \topcaption{
       Summary of systematic uncertainties included in the analysis. The value given for each rate uncertainty is the maximum value before the fit is performed. Uncertainties in the shape of the distributions are instead labeled as such. 
    }
    \label{tab:systematics}
    \begin{tabular}{l c c c}
        \hline
         Source &  Uncertainty \\
        \hline
        Integrated luminosity
          & 0.6--2\% \\

        Pileup
          & 1\% \\

        L1 prefiring
          & 1\% \\

        \ptmiss trigger efficiency
          & 1--2\% \\

        Single-electron trigger efficiency
          & 1\% \\

        Muon isolation efficiency
          & 1\% \\

        Muon identification efficiency
          & 1\% \\

        Electron reconstruction efficiency
          & 1\% \\

        Electron identification efficiency
          & 2--3\% \\

        Tau veto
          & 3\% \\   

        Hadronic recoil \recoil 
          & 5\% \\

        Jet energy corrections
          & 4\% \\

        \textsc{DeepJet} efficiency
          & Shape \\

        \textsc{DeepAK15} efficiency
          & Shape \\

        $\PV\PV$ cross section
          & 20\% \\

        $\PH {\to} \bbbar$ cross section
          & 20\% \\

        \DYjets cross section
          & 20\% \\

        Single \PQt cross section
          & 10\% \\

        \ttbar cross section
          & 10\% \\

        QCD-\ptmiss normalization
          & 100\% \\

        QCD-electron normalization
          & 100\% \\

        QCD-muon normalization
          & 100\% \\

        Higher-order corrections
          & Shape \\

        Bin-by-bin event counts
          & Shape \\
        \hline
    \end{tabular}
\end{table}

\section{Results and interpretation}

\begin{figure*}[!htbp]
  \centering
  \includegraphics[width=1.0\textwidth]{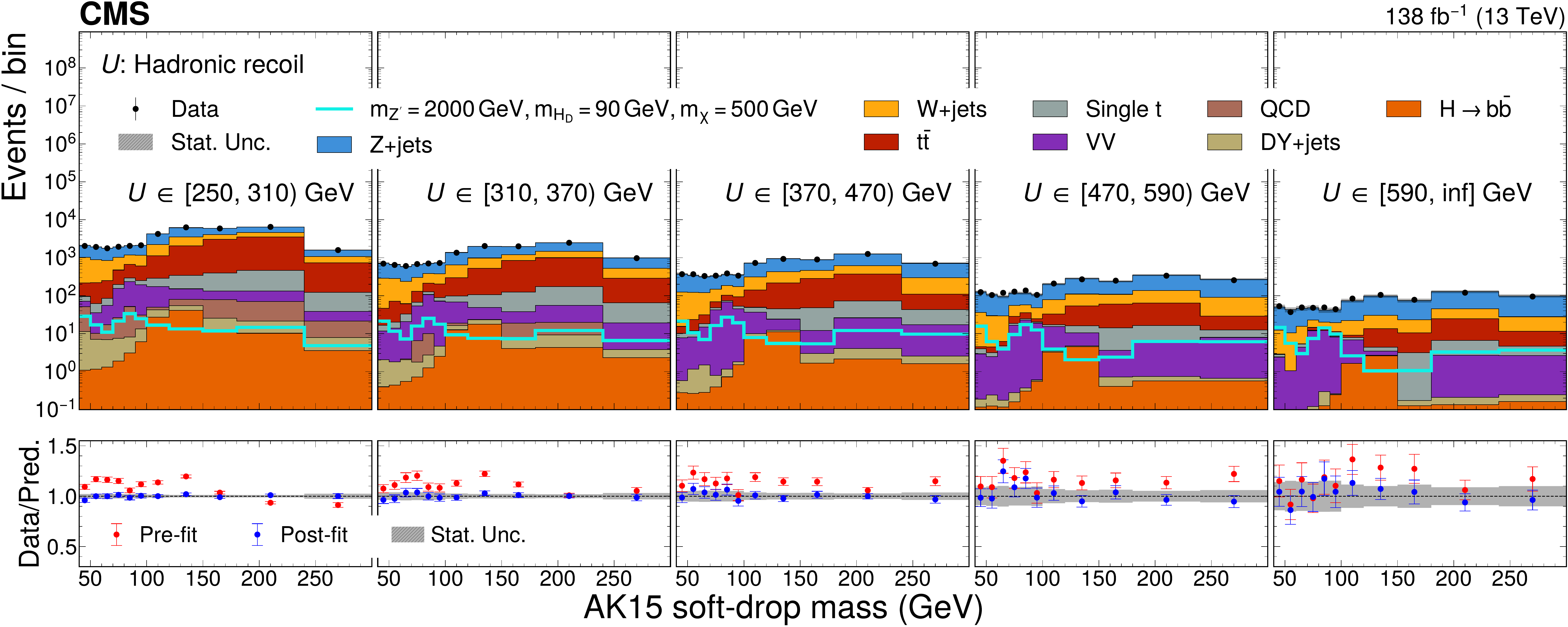}\\
  \caption{Post-fit \mSD distributions in bins of of the hadronic recoil \recoil. The upper panels present stacked post-fit predictions for the backgrounds superimposed on the data, with an example signal hypothesis overlaid. The lower panels show in red (blue) the pre-fit (post-fit) values of the ratio between the observed data and the predicted SM background, with their associated uncertainties. The gray area represents the total statistical post-fit uncertainty.}
  \label{sr}
\end{figure*}

The maximum likelihood fit is performed by combining the SR and CRs as well as the data sets corresponding to all the data-taking years. Contributions from both signal and background processes are included in the fit. The \recoil-vs.-$\mSD$ distributions in the SR before and after the fit (``pre-fit'' and ``post-fit'') for all three years combined are shown in Fig.~\ref{sr}.

No significant excess is observed above the expected background. Exclusion limits on the signal strength $\mu$ are presented for all signal hypotheses, using all data sets and analysis regions, and calculated with the \CLs criterion~\cite{CLS1,CLS2}. An asymptotic approximation of the distribution of the profile likelihood test statistic~\cite{Cowan:2010js} is employed.

\begin{figure*}[!htbp]
    \centering
      \includegraphics[width=0.49\textwidth]{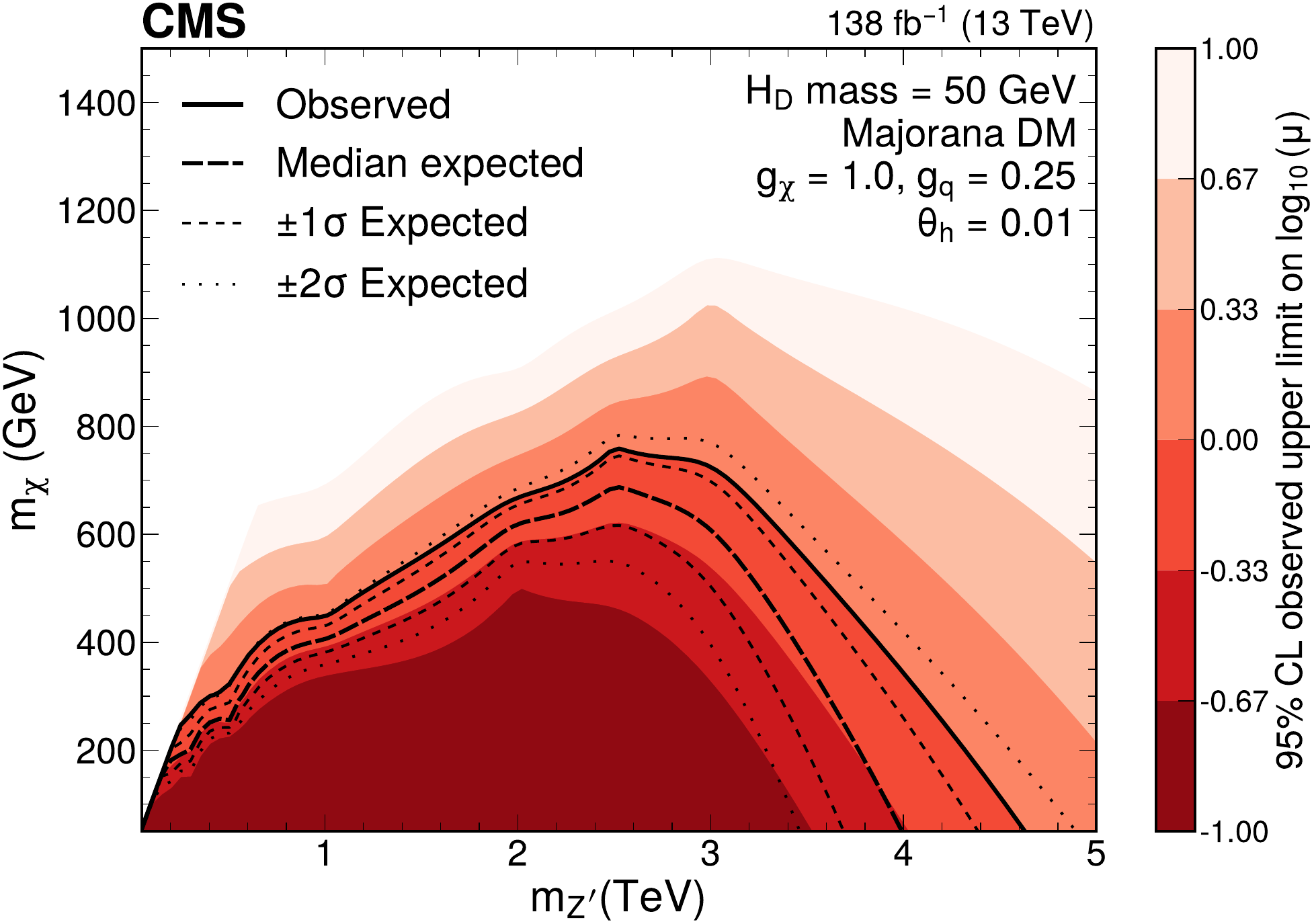}~~~\includegraphics[width=0.49\textwidth]{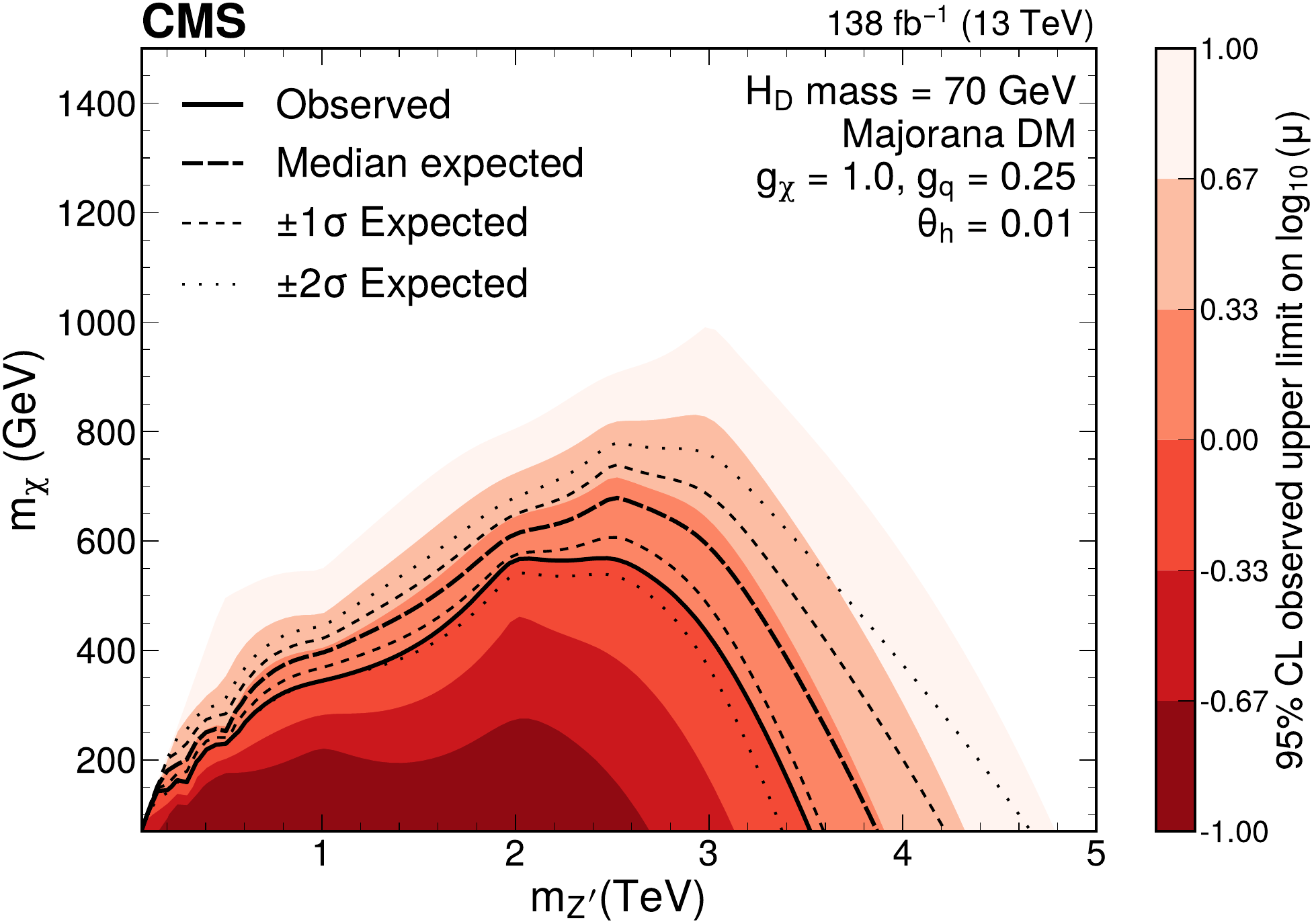}\\
      \includegraphics[width=0.49\textwidth]{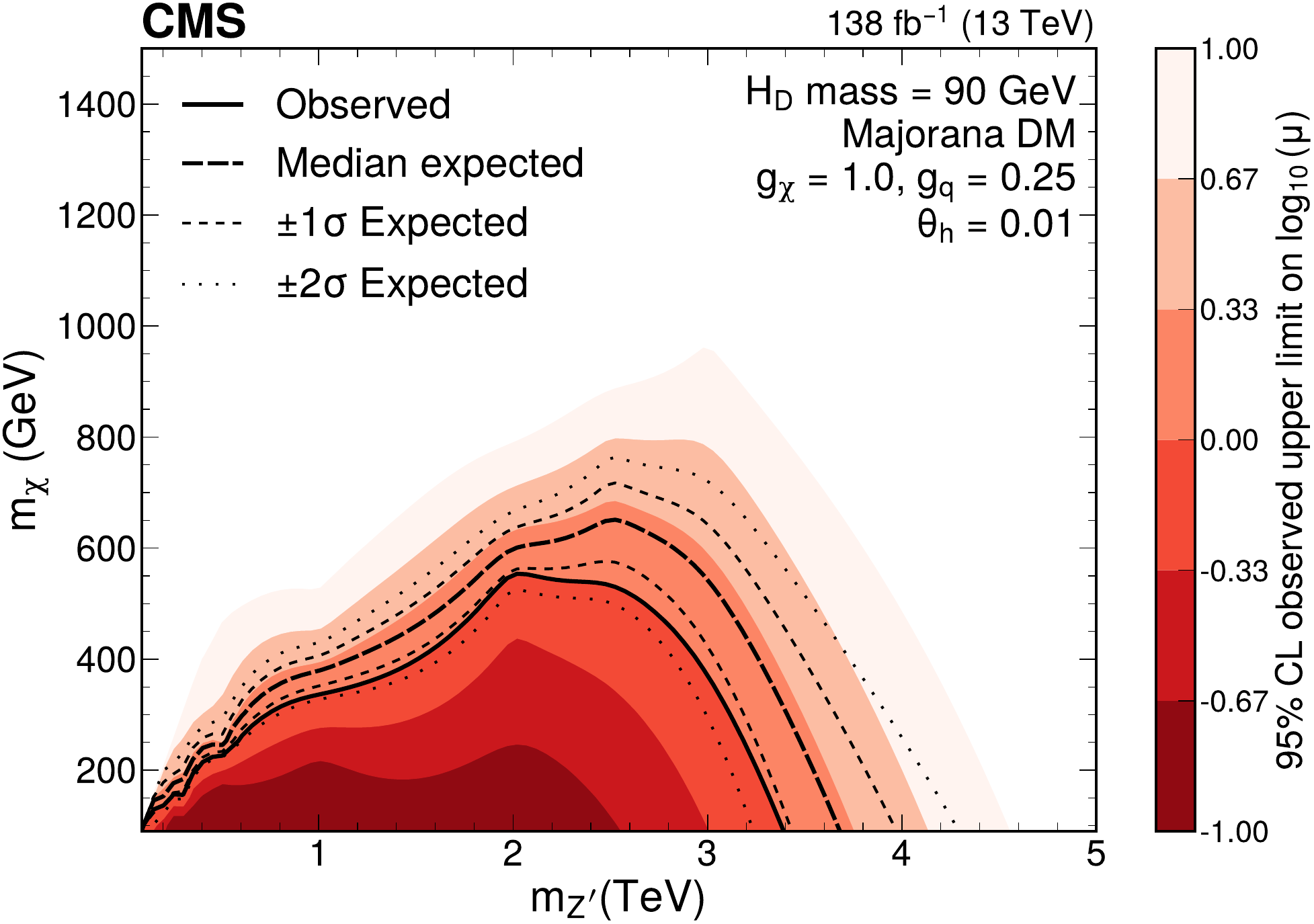}~~~\includegraphics[width=0.49\textwidth]{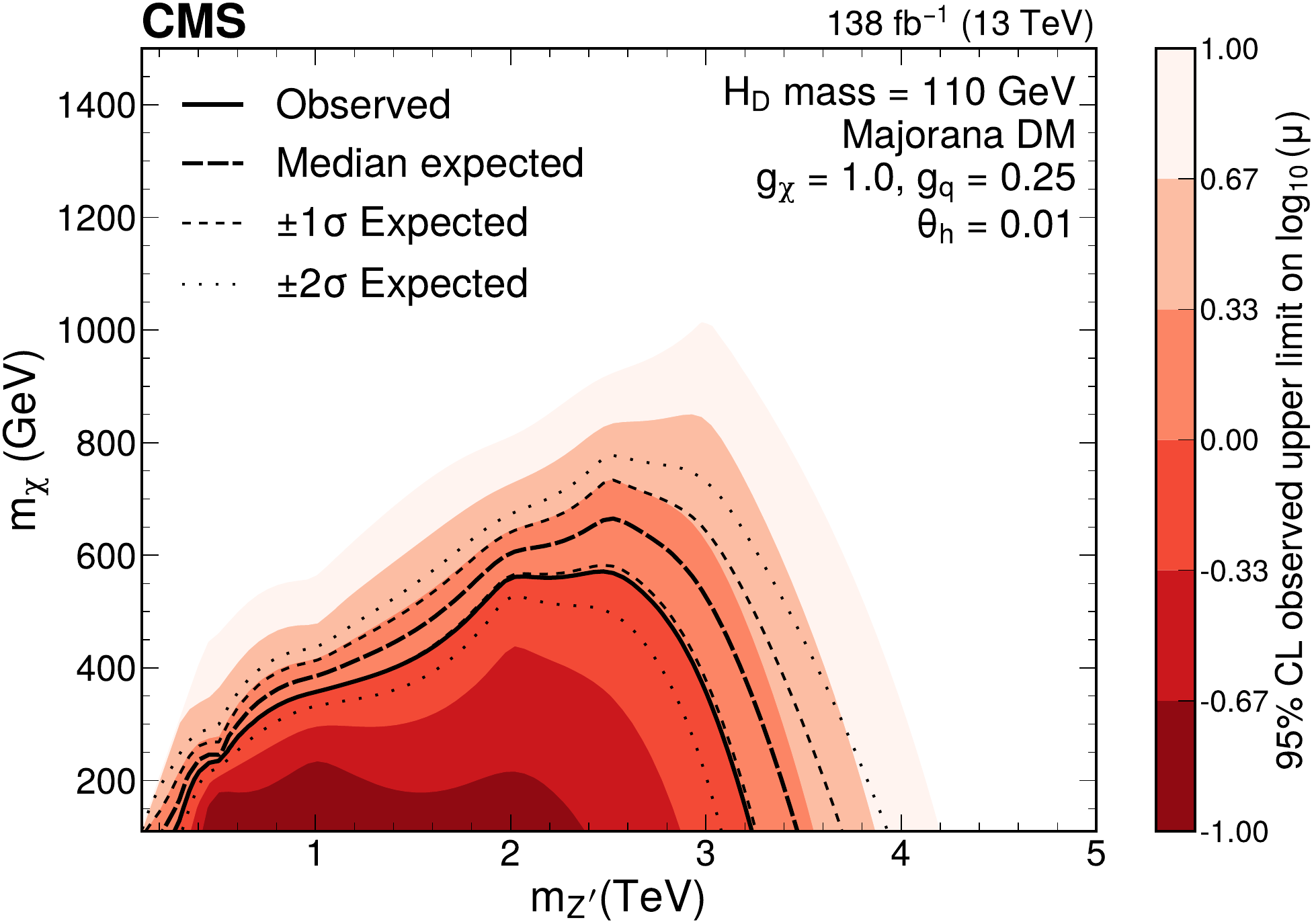}\\
      \includegraphics[width=0.49\textwidth]{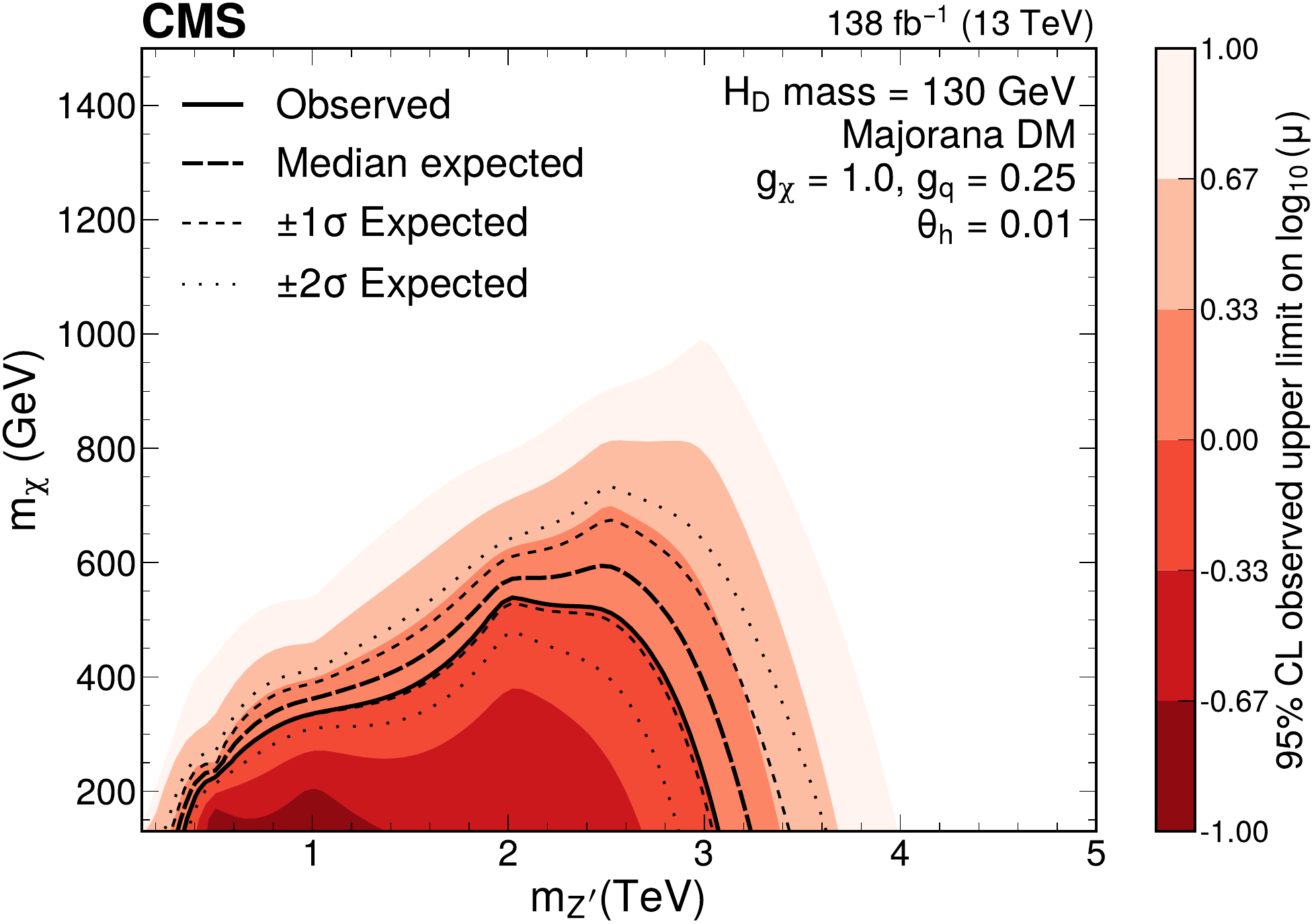}~~~\includegraphics[width=0.49\textwidth]{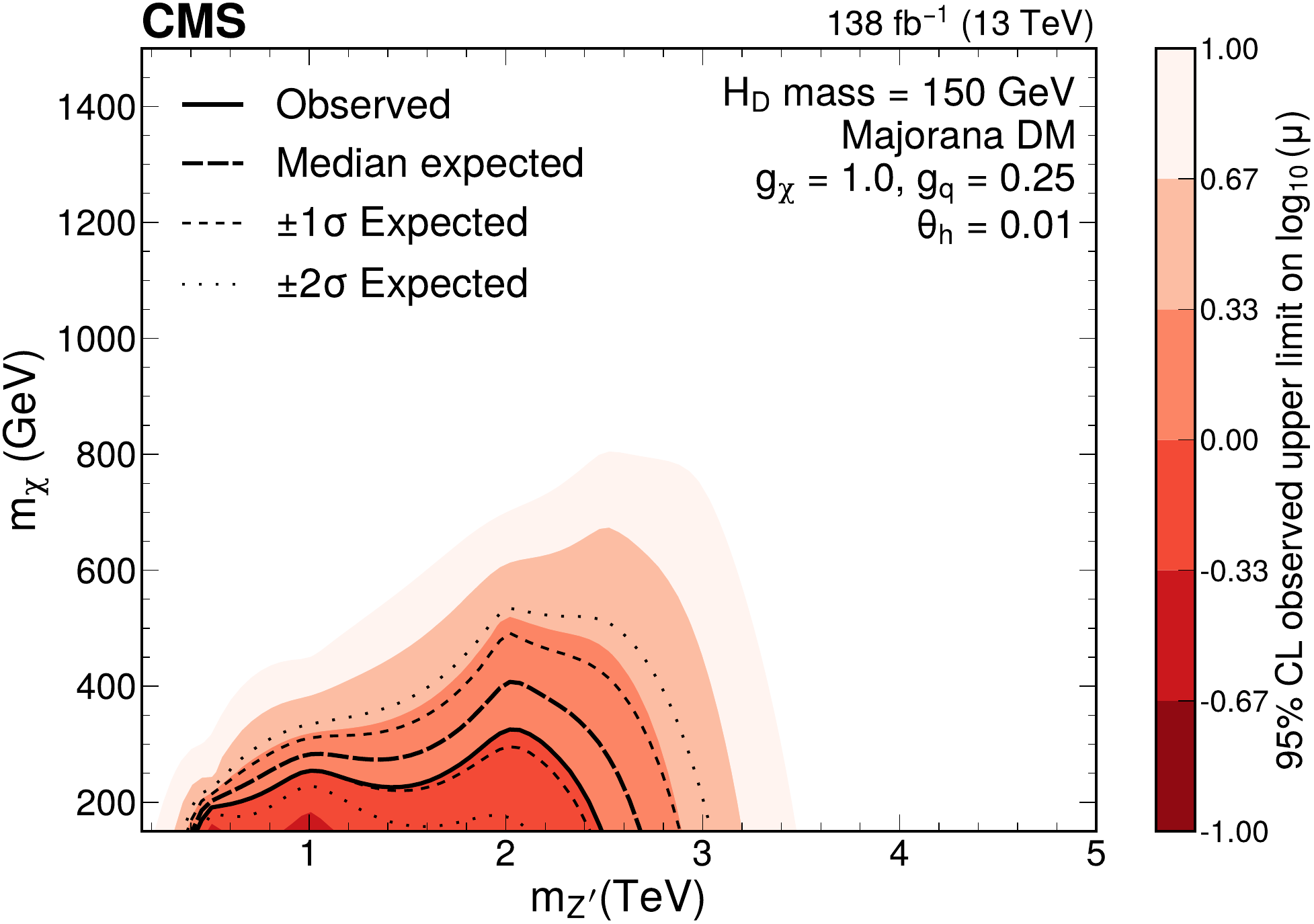}
      \caption{Expected and observed exclusion limits at 95\% \CL on the signal strength $\mu=\sigma/\sigma_\text{theory}$ as a function of the mediator mass \mzp and dark matter mass $\mchi$ for six dark Higgs boson masses ($\darkHiggs$ mass).  Only scenarios with DM heavier than the \darkHiggs boson are considered. The black solid line indicates the observed exclusion boundary corresponding to $\mu=1$. The black dashed and dotted lines represent the expected exclusion and the 68 and 95\% \CL intervals around the expected boundary, respectively. Parameter combinations corresponding to larger values of $\mu$ are excluded.}
      \label{fig:mhs}
\end{figure*}

Two-dimensional exclusion limits are calculated in the parameter space of the DM and mediator masses, \mchi and \mzp, constrained by the fact that only scenarios in which the DM particle is more massive than the \darkHiggs boson are considered. The resulting exclusion limits at 95\% confidence level (\CL) on $\mu$ are shown in Fig.~\ref{fig:mhs} for different hypotheses of $m_{\darkHiggs}$. In the plots, darker shades correspond to smaller upper limits, i.e., more stringent constraints. The solid black line represents the observed 95\% \CL exclusion contour, while the dashed and dotted lines indicate the median expected exclusion and its 68\% and 95\% confidence intervals, respectively. The parameter space inside the solid black boundary is excluded at 95\% \CL under the model assumptions. Values of the mediator mass up to 4.5 (2.5)\TeV are excluded for an $m_{\darkHiggs}$ of 50 (150)\GeV. The ATLAS Collaboration recently published their results on the same final state~\cite{ATLAS:2024ypx}. They present limits for $m_{\darkHiggs}$ in the range of 30--150 GeV using 13\TeV proton-proton collision data corresponding to an integrated luminosity of 140\fbinv. They observe no excess over the expected background contributions and exclude mediator masses up to 3.4\TeV for the benchmark model with \gchi = 1.0, \gq = 0.25, and $\theta_{\mathrm{h}} = 0.01$. When compared to the results from the CMS experiment described in this paper for an \mchi hypothesis of 200\GeV, CMS excludes larger values of \mzp for all the $m_{\darkHiggs}$ hypotheses considered in this search (50 to 150\GeV). 

\section{Summary}
A search for dark matter (DM) produced in association with a dark Higgs boson (\darkHiggs) in events with a resonant bottom quark-antiquark pair and large missing transverse momentum has been presented. Hypotheses for \darkHiggs bosons with masses between 50--150\GeV are tested with this search. Only scenarios in which the DM particle is more massive than the \darkHiggs boson are considered. A data set of proton-proton collisions at a center-of-mass energy of 13\TeV corresponding to an integrated luminosity of 138\fbinv is analyzed. A joint maximum likelihood fit spanning a set of one signal region and multiple control regions is used to constrain the standard model background contributions to the data and to extract a possible signal. No significant excess is observed above the background.

The result is interpreted in terms of exclusion limits at 95\% confidence level on the parameters of the model. Values of the mediator mass up to 4.5 (2.5)\TeV are excluded for an \darkHiggs boson of 50 (150)\GeV, assuming couplings of $\gq = 0.25$ between the mediator and quarks, of $\gchi = 1.0$ between the mediator and the DM particles, and a mixing angle between the \darkHiggs boson and the standard model Higgs boson of 0.01. 

\begin{acknowledgments}
We congratulate our colleagues in the CERN accelerator departments for the excellent performance of the LHC and thank the technical and administrative staffs at CERN and at other CMS institutes for their contributions to the success of the CMS effort. In addition, we gratefully acknowledge the computing centers and personnel of the Worldwide LHC Computing Grid and other centers for delivering so effectively the computing infrastructure essential to our analyses. Finally, we acknowledge the enduring support for the construction and operation of the LHC, the CMS detector, and the supporting computing infrastructure provided by the following funding agencies: SC (Armenia), BMBWF and FWF (Austria); FNRS and FWO (Belgium); CNPq, CAPES, FAPERJ, FAPERGS, and FAPESP (Brazil); MES and BNSF (Bulgaria); CERN; CAS, MoST, and NSFC (China); MINCIENCIAS (Colombia); MSES and CSF (Croatia); RIF (Cyprus); SENESCYT (Ecuador); ERC PRG and PSG, TARISTU24-TK10 and MoER TK202 (Estonia); Academy of Finland, MEC, and HIP (Finland); CEA and CNRS/IN2P3 (France); SRNSF (Georgia); BMFTR, DFG, and HGF (Germany); GSRI (Greece); MATE and NKFIH (Hungary); DAE and DST (India); IPM (Iran); SFI (Ireland); INFN (Italy); MSIT and NRF (Republic of Korea); MES (Latvia); LMTLT (Lithuania); MOE and UM (Malaysia); BUAP, CINVESTAV, CONACYT, LNS, SEP, and UASLP-FAI (Mexico); MOS (Montenegro); MBIE (New Zealand); PAEC (Pakistan); MSHE, NSC, and NAWA (Poland); FCT (Portugal); MESTD (Serbia); MICIU/AEI and PCTI (Spain); MOSTR (Sri Lanka); Swiss Funding Agencies (Switzerland); MST (Taipei); MHESI (Thailand); TUBITAK and TENMAK (T\"{u}rkiye); NASU (Ukraine); STFC (United Kingdom); DOE and NSF (USA).

\hyphenation{Rachada-pisek} Individuals have received support from the Marie-Curie program and the European Research Council and Horizon 2020 Grant, contract Nos.\ 675440, 724704, 752730, 758316, 765710, 824093, 101115353, 101002207, 101001205, and COST Action CA16108 (European Union); the Leventis Foundation; the Alfred P.\ Sloan Foundation; the Alexander von Humboldt Foundation; the Science Committee, project no. 22rl-037 (Armenia); the Fonds pour la Formation \`a la Recherche dans l'Industrie et dans l'Agriculture (FRIA) and Fonds voor Wetenschappelijk Onderzoek contract No. 1228724N (Belgium); the Beijing Municipal Science \& Technology Commission, No. Z191100007219010, the Fundamental Research Funds for the Central Universities, the Ministry of Science and Technology of China under Grant No. 2023YFA1605804, the Natural Science Foundation of China under Grant No. 12535004, and USTC Research Funds of the Double First-Class Initiative No.\ YD2030002017 (China); the Ministry of Education, Youth and Sports (MEYS) of the Czech Republic; the Shota Rustaveli National Science Foundation, grant FR-22-985 (Georgia); the Deutsche Forschungsgemeinschaft (DFG), among others, under Germany's Excellence Strategy -- EXC 2121 ``Quantum Universe" -- 390833306, and under project number 400140256 - GRK2497; the Hellenic Foundation for Research and Innovation (HFRI), Project Number 2288 (Greece); the Hungarian Academy of Sciences, the New National Excellence Program - \'UNKP, the NKFIH research grants K 131991, K 138136, K 143460, K 143477, K 147557, K 146913, K 146914, K 147048, TKP2021-NKTA-64, and 2025-1.1.5-NEMZ\_KI-2025-00004, and MATE KKP and KKPCs Research Excellence and Flagship Research Groups grants (Hungary); the Council of Science and Industrial Research, India; ICSC -- National Research Center for High Performance Computing, Big Data and Quantum Computing, FAIR -- Future Artificial Intelligence Research, and CUP I53D23001070006 (Mission 4 Component 1), funded by the NextGenerationEU program, the Italian Ministry of University and Research (MUR) under Bando PRIN 2022 -- CUP I53C24002390006, PRIN PRIMULA 2022RBYK7T (Italy); the Latvian Council of Science; the Ministry of Science and Higher Education, project no. 2022/WK/14, and the National Science Center, contracts Opus 2021/41/B/ST2/01369, 2021/43/B/ST2/01552, 2023/49/B/ST2/03273, and the NAWA contract BPN/PPO/2021/1/00011 (Poland); the Funda\c{c}\~ao para a Ci\^encia e a Tecnologia (Portugal); the National Priorities Research Program by Qatar National Research Fund; MICIU/AEI/10.13039/501100011033, ERDF/EU, ``European Union NextGenerationEU/PRTR", projects PID2022-142604OB-C21, PID2022-139519OB-C21, PID2023-147706NB-I00, PID2023-148896NB-I00, PID2023-146983NB-I00, PID2023-147115NB-I00, PID2023-148418NB-C41, PID2023-148418NB-C42, PID2023-148418NB-C43, PID2023-148418NB-C44, PID2024-158190NB-C22, RYC2021-033305-I, RYC2024-048719-I, CNS2023-144781, CNS2024-154769 and Plan de Ciencia, Tecnolog{\'i}a e Innovaci{\'o}n de Asturias, Spain; the Chulalongkorn Academic into Its 2nd Century Project Advancement Project, the National Science, Research and Innovation Fund program IND\_FF\_68\_369\_2300\_097, and the Program Management Unit for Human Resources \& Institutional Development, Research and Innovation, grant B39G680009 (Thailand); the Eric \& Wendy Schmidt Fund for Strategic Innovation through the CERN Next Generation Triggers project under grant agreement number SIF-2023-004; the Kavli Foundation; the Nvidia Corporation; the SuperMicro Corporation; the Welch Foundation, contract C-1845; and the Weston Havens Foundation (USA).
\end{acknowledgments}\section*{Data availability} Release and preservation of data used by the CMS Collaboration as the basis for publications is guided by the  \href{https://doi.org/10.7483/OPENDATA.CMS.1BNU.8V1W}{CMS data preservation, re-use and open access policy}.
\bibliography{auto_generated}
\cleardoublepage \appendix\section{The CMS Collaboration \label{app:collab}}\begin{sloppypar}\hyphenpenalty=5000\widowpenalty=500\clubpenalty=5000\input{SUS-23-013-public-authorlist.tex}\end{sloppypar}
\end{document}

%% file: SUS-23-013-public-authorlist.tex
\cmsinstitute{Yerevan Physics Institute, Yerevan, Armenia}
{\tolerance=6000
A.~Hayrapetyan, V.~Makarenko\cmsorcid{0000-0002-8406-8605}, A.~Tumasyan\cmsAuthorMark{1}\cmsorcid{0009-0000-0684-6742}
\par}
\cmsinstitute{Institut f\"{u}r Hochenergiephysik, Vienna, Austria}
{\tolerance=6000
W.~Adam\cmsorcid{0000-0001-9099-4341}, L.~Benato\cmsorcid{0000-0001-5135-7489}, T.~Bergauer\cmsorcid{0000-0002-5786-0293}, M.~Dragicevic\cmsorcid{0000-0003-1967-6783}, C.~Giordano\cmsorcid{0000-0001-6317-2481}, P.S.~Hussain\cmsorcid{0000-0002-4825-5278}, M.~Jeitler\cmsAuthorMark{2}\cmsorcid{0000-0002-5141-9560}, N.~Krammer\cmsorcid{0000-0002-0548-0985}, A.~Li\cmsorcid{0000-0002-4547-116X}, D.~Liko\cmsorcid{0000-0002-3380-473X}, M.~Matthewman, J.~Schieck\cmsAuthorMark{2}\cmsorcid{0000-0002-1058-8093}, R.~Sch\"{o}fbeck\cmsAuthorMark{2}\cmsorcid{0000-0002-2332-8784}, M.~Shooshtari\cmsorcid{0009-0004-8882-4887}, M.~Sonawane\cmsorcid{0000-0003-0510-7010}, W.~Waltenberger\cmsorcid{0000-0002-6215-7228}, C.-E.~Wulz\cmsAuthorMark{2}\cmsorcid{0000-0001-9226-5812}
% ...
\par}
\vskip\cmsinstskip
\dag:~Deceased\\
$^{1}$Also at Yerevan State University, Yerevan, Armenia\\
$^{2}$Also at TU Wien, Vienna, Austria\\
% ...


%% file: auto_generated.bib
@ARTICLE{Arkani_Hamed_2009,
	AUTHOR=	"Arkani-Hamed, Nima and Finkbeiner, Douglas P. and Slatyer, Tracy R. and Weiner, Neal",
	TITLE=	"A Theory of Dark Matter",
	EPRINT=	"0810.0713",
	ARCHIVEPREFIX=	"arXiv",
	PRIMARYCLASS=	"hep-ph",
	DOI=	"10.1103/PhysRevD.79.015014",
	JOURNAL=	"Phys. Rev. D",
	VOLUME=	"79",
	PAGES=	"015014",
	YEAR=	"2009",
}

@ARTICLE{1937ApJ,
	AUTHOR=	"Zwicky, F.",
	TITLE=	"On the Masses of Nebulae and of Clusters of Nebulae",
	JOURNAL=	"Astrophys. J.",
	YEAR=	"1937",
	VOLUME=	"86",
	PAGES=	"217",
	DOI=	"10.1086/143864",
}

@ARTICLE{Bertone:2004pz,
	AUTHOR=	"Bertone, Gianfranco and Hooper, Dan and Silk, Joseph",
	TITLE=	"Particle dark matter: Evidence, candidates and constraints",
	EPRINT=	"hep-ph/0404175",
	ARCHIVEPREFIX=	"arXiv",
	REPORTNUMBER=	"FERMILAB-PUB-04-047-A",
	DOI=	"10.1016/j.physrep.2004.08.031",
	JOURNAL=	"Phys. Rept.",
	VOLUME=	"405",
	PAGES=	"279",
	YEAR=	"2005",
}

@ARTICLE{Abercrombie_2020,
	AUTHOR=	"Abercrombie, Daniel and others",
	EDITOR=	"Boveia, Antonio and Doglioni, Caterina and Lowette, Steven and Malik, Sarah and Mrenna, Stephen",
	TITLE=	"Dark Matter benchmark models for early {LHC} {Run-2} Searches: Report of the {ATLAS/CMS} Dark Matter Forum",
	EPRINT=	"1507.00966",
	ARCHIVEPREFIX=	"arXiv",
	PRIMARYCLASS=	"hep-ex",
	REPORTNUMBER=	"FERMILAB-PUB-15-282-CD",
	DOI=	"10.1016/j.dark.2019.100371",
	JOURNAL=	"Phys. Dark Univ.",
	VOLUME=	"27",
	PAGES=	"100371",
	YEAR=	"2020",
}

@ARTICLE{CMS:2021far,
	AUTHOR=	"{CMS Collaboration}",
	TITLE=	"Search for new particles in events with energetic jets and large missing transverse momentum in proton-proton collisions at $\sqrt{s}$ = 13 {TeV}",
	EPRINT=	"2107.13021",
	ARCHIVEPREFIX=	"arXiv",
	PRIMARYCLASS=	"hep-ex",
	REPORTNUMBER=	"CMS-EXO-20-004, CERN-EP-2021-136",
	DOI=	"10.1007/JHEP11(2021)153",
	JOURNAL=	"JHEP",
	VOLUME=	"11",
	PAGES=	"153",
	YEAR=	"2021",
}

@ARTICLE{PhysRevD.103.112006,
	AUTHOR=	"{ATLAS Collaboration}",
	TITLE=	"Search for new phenomena in events with an energetic jet and missing transverse momentum in {$\Pp\Pp$} collisions at $\sqrt{s}$ =13 {TeV} with the {ATLAS} detector",
	EPRINT=	"2102.10874",
	ARCHIVEPREFIX=	"arXiv",
	PRIMARYCLASS=	"hep-ex",
	REPORTNUMBER=	"CERN-EP-2020-238",
	DOI=	"10.1103/PhysRevD.103.112006",
	JOURNAL=	"Phys. Rev. D",
	VOLUME=	"103",
	PAGES=	"112006",
	YEAR=	"2021",
}

@ARTICLE{Duerr:2017uap,
	AUTHOR=	"Duerr, Michael and Grohsjean, Alexander and Kahlhoefer, Felix and Penning, Bjoern and Schmidt-Hoberg, Kai and Schwanenberger, Christian",
	TITLE=	"Hunting the dark {Higgs}",
	EPRINT=	"1701.08780",
	ARCHIVEPREFIX=	"arXiv",
	PRIMARYCLASS=	"hep-ph",
	REPORTNUMBER=	"DESY-17-016",
	DOI=	"10.1007/JHEP04(2017)143",
	JOURNAL=	"JHEP",
	VOLUME=	"04",
	PAGES=	"143",
	YEAR=	"2017",
}

@ARTICLE{Bell:2016fqf,
	AUTHOR=	"Bell, Nicole F. and Cai, Yi and Leane, Rebecca K.",
	TITLE=	"Dark Forces in the Sky: Signals from {\PZpr} and the Dark {Higgs}",
	EPRINT=	"1605.09382",
	ARCHIVEPREFIX=	"arXiv",
	PRIMARYCLASS=	"hep-ph",
	DOI=	"10.1088/1475-7516/2016/08/001",
	JOURNAL=	"JCAP",
	VOLUME=	"08",
	PAGES=	"001",
	YEAR=	"2016",
}

@ARTICLE{Kahlhoefer:2015bea,
	AUTHOR=	"Kahlhoefer, Felix and Schmidt-Hoberg, Kai and Schwetz, Thomas and Vogl, Stefan",
	TITLE=	"Implications of unitarity and gauge invariance for simplified dark matter models",
	JOURNAL=	"JHEP",
	VOLUME=	"02",
	YEAR=	"2016",
	PAGES=	"016",
	DOI=	"10.1007/JHEP02(2016)016",
	EPRINT=	"1510.02110",
	ARCHIVEPREFIX=	"arXiv",
	PRIMARYCLASS=	"hep-ph",
	REPORTNUMBER=	"DESY-15-182",
	SLACCITATION=	"%%CITATION = ARXIV:1510.02110;%%",
}

@ARTICLE{Bell:2016uhg,
	AUTHOR=	"Bell, Nicole F. and Cai, Yi and Leane, Rebecca K.",
	TITLE=	"Impact of mass generation for spin-1 mediator simplified models",
	EPRINT=	"1610.03063",
	ARCHIVEPREFIX=	"arXiv",
	PRIMARYCLASS=	"hep-ph",
	DOI=	"10.1088/1475-7516/2017/01/039",
	JOURNAL=	"JCAP",
	VOLUME=	"01",
	PAGES=	"039",
	YEAR=	"2017",
}

@ARTICLE{Albert:2017onk,
	AUTHOR=	"{LHC Dark Matter Working Group}",
	ARCHIVEPREFIX=	"arXiv",
	DOI=	"10.1016/j.dark.2019.100377",
	EPRINT=	"1703.05703",
	JOURNAL=	"Phys. Dark Univ.",
	PAGES=	"100377",
	PRIMARYCLASS=	"hep-ex",
	REPORTNUMBER=	"CERN-LPCC-2017-01",
	TITLE=	"Recommendations of the {LHC} {Dark Matter Working Group}: Comparing {LHC} searches for dark matter mediators in visible and invisible decay channels and calculations of the thermal relic density",
	VOLUME=	"26",
	YEAR=	"2019",
}

@ARTICLE{Frandsen:2012rk,
	AUTHOR=	"Frandsen, Mads T. and Kahlhoefer, Felix and Preston, Anthony and Sarkar, Subir and Schmidt-Hoberg, Kai",
	TITLE=	"{LHC} and {Tevatron} Bounds on the Dark Matter Direct Detection Cross-Section for Vector Mediators",
	JOURNAL=	"JHEP",
	VOLUME=	"07",
	YEAR=	"2012",
	PAGES=	"123",
	DOI=	"10.1007/JHEP07(2012)123",
	EPRINT=	"1204.3839",
	ARCHIVEPREFIX=	"arXiv",
	PRIMARYCLASS=	"hep-ph",
	REPORTNUMBER=	"OUTP-12-05P",
	SLACCITATION=	"%%CITATION = ARXIV:1204.3839;%%",
}

@ARTICLE{CMS:2022dwd,
	AUTHOR=	"{CMS Collaboration}",
	TITLE=	"A portrait of the {Higgs} boson by the {CMS} experiment ten years after the discovery",
	EPRINT=	"2207.00043",
	ARCHIVEPREFIX=	"arXiv",
	PRIMARYCLASS=	"hep-ex",
	REPORTNUMBER=	"CMS-HIG-22-001, CERN-EP-2022-039",
	DOI=	"10.1038/s41586-022-04892-x",
	JOURNAL=	"Nature",
	VOLUME=	"607",
	PAGES=	"60",
	YEAR=	"2022",
	NOTE=	"[Corrigendum: \DOI{10.1038/s41586-023-06164-8}]",
}

@ARTICLE{ATLAS:2022bzt,
	AUTHOR=	"{ATLAS Collaboration}",
	TITLE=	"Search for dark matter produced in association with a dark {Higgs} boson decaying into {$\PW^+\PW^-$} in the one-lepton final state at $\sqrt{s}$ = 13 {TeV} using 139 fb$^{-1}$ of {$\Pp\Pp$} collisions recorded with the {ATLAS} detector",
	EPRINT=	"2211.07175",
	ARCHIVEPREFIX=	"arXiv",
	PRIMARYCLASS=	"hep-ex",
	REPORTNUMBER=	"CERN-EP-2022-147",
	DOI=	"10.1007/JHEP07(2023)116",
	JOURNAL=	"JHEP",
	VOLUME=	"07",
	PAGES=	"116",
	YEAR=	"2023",
}

@ARTICLE{CMS:2023dof,
	AUTHOR=	"{CMS Collaboration}",
	TITLE=	"Search for dark matter particles in {$\PW^+\PW^-$} events with transverse momentum imbalance in proton-proton collisions at $\sqrt{s}$ = 13 {TeV}",
	EPRINT=	"2310.12229",
	ARCHIVEPREFIX=	"arXiv",
	PRIMARYCLASS=	"hep-ex",
	REPORTNUMBER=	"CMS-EXO-21-012, CERN-EP-2023-216",
	DOI=	"10.1007/JHEP03(2024)134",
	JOURNAL=	"JHEP",
	VOLUME=	"03",
	PAGES=	"134",
	YEAR=	"2024",
}

@ARTICLE{ATLAS:2024ypx,
	AUTHOR=	"{ATLAS Collaboration}",
	TITLE=	"Search for Dark Matter Produced in Association with a Dark {Higgs} Boson in the {$\PQb\PAQb$} Final State Using {$\Pp\Pp$} Collisions at $\sqrt{s}$ = 13 {TeV} with the {ATLAS} Detector",
	EPRINT=	"2407.10549",
	ARCHIVEPREFIX=	"arXiv",
	PRIMARYCLASS=	"hep-ex",
	REPORTNUMBER=	"CERN-EP-2024-170",
	DOI=	"10.1103/PhysRevLett.134.121801",
	JOURNAL=	"Phys. Rev. Lett.",
	VOLUME=	"134",
	PAGES=	"121801",
	YEAR=	"2025",
}

@MISC{hepdata,
	TITLE=	"{HEPD}ata record for this analysis",
	DOI=	"10.17182/hepdata.170081",
	YEAR=	"2026",
}

@ARTICLE{Chatrchyan:2008zzk,
	AUTHOR=	"{CMS Collaboration}",
	TITLE=	"The {CMS} experiment at the {CERN} {LHC}",
	JOURNAL=	"JINST",
	VOLUME=	"3",
	YEAR=	"2008",
	PAGES=	"S08004",
	DOI=	"10.1088/1748-0221/3/08/S08004",
	SLACCITATION=	"%%CITATION = JINST,3,S08004;%%",
}

@ARTICLE{CMS:2023gfb,
	AUTHOR=	"{CMS Collaboration}",
	TITLE=	"Development of the {CMS} detector for the {CERN LHC Run 3}",
	DOI=	"10.1088/1748-0221/19/05/P05064",
	EPRINT=	"2309.05466",
	JOURNAL=	"JINST",
	VOLUME=	"19",
	PAGES=	"P05064",
	YEAR=	"2024",
}

@ARTICLE{TRK-11-001,
	AUTHOR=	"{CMS Collaboration}",
	TITLE=	"Description and performance of track and primary-vertex reconstruction with the {CMS} tracker",
	JOURNAL=	"JINST",
	VOLUME=	"9",
	PAGES=	"P10009",
	DOI=	"10.1088/1748-0221/9/10/P10009",
	YEAR=	"2014",
	EPRINT=	"1405.6569",
	ARCHIVEPREFIX=	"arXiv",
	PRIMARYCLASS=	"physics.ins-det",
	REPORTNUMBER=	"CMS-TRK-11-001, CERN-PH-EP-2014-070",
	SLACCITATION=	"%%CITATION = ARXIV:1405.6569;%%",
}

@ARTICLE{Phase1Pixel,
	AUTHOR=	"Adam, W. and others",
	COLLABORATION=	"Tracker Group in the CMS",
	TITLE=	"The {CMS} {Phase-1} Pixel Detector Upgrade",
	EPRINT=	"2012.14304",
	ARCHIVEPREFIX=	"arXiv",
	PRIMARYCLASS=	"physics.ins-det",
	DOI=	"10.1088/1748-0221/16/02/P02027",
	JOURNAL=	"JINST",
	VOLUME=	"16",
	PAGES=	"P02027",
	YEAR=	"2021",
}

@TECHREPORT{DP-2020-049,
	TITLE=	"Track impact parameter resolution for the full pseudo rapidity coverage in the 2017 dataset with the {CMS} {Phase-1} Pixel detector",
	URL=	"https://cds.cern.ch/record/2743740",
	TYPE=	"CMS Detector Performance Note",
	NUMBER=	"CMS-DP-2020-049",
	YEAR=	"2020",
	AUTHOR=	"{CMS Collaboration}",
}

@TECHREPORT{CMS-TDR-15-02,
	AUTHOR=	"{CMS Collaboration}",
	TITLE=	"Technical proposal for the {Phase-II} upgrade of the {Compact Muon Solenoid}",
	URL=	"http://cds.cern.ch/record/2020886",
	TYPE=	"CMS Technical Proposal",
	NUMBER=	"CERN-LHCC-2015-010, CMS-TDR-15-02",
	YEAR=	"2015",
}

@ARTICLE{Sirunyan:2020zal,
	AUTHOR=	"{CMS Collaboration}",
	TITLE=	"Performance of the {CMS} {Level-1} trigger in proton-proton collisions at $\sqrt{s} = 13$\,{TeV}",
	JOURNAL=	"JINST",
	VOLUME=	"15",
	PAGES=	"P10017",
	YEAR=	"2020",
	DOI=	"10.1088/1748-0221/15/10/P10017",
	EPRINT=	"2006.10165",
	ARCHIVEPREFIX=	"arXiv",
	PRIMARYCLASS=	"hep-ex",
	REPORTNUMBER=	"CMS-TRG-17-001, CERN-EP-2020-065",
}

@ARTICLE{Khachatryan:2016bia,
	AUTHOR=	"{CMS Collaboration}",
	TITLE=	"The {CMS} trigger system",
	JOURNAL=	"JINST",
	VOLUME=	"12",
	PAGES=	"P01020",
	DOI=	"10.1088/1748-0221/12/01/P01020",
	YEAR=	"2017",
	EPRINT=	"1609.02366",
	ARCHIVEPREFIX=	"arXiv",
	PRIMARYCLASS=	"physics.ins-det",
	REPORTNUMBER=	"CMS-TRG-12-001, CERN-EP-2016-160",
	SLACCITATION=	"%%CITATION = ARXIV:1609.02366;%%",
}

@ARTICLE{CMS:2024aqx,
	AUTHOR=	"{CMS Collaboration}",
	TITLE=	"Performance of the {CMS} high-level trigger during {LHC Run 2}",
	EPRINT=	"2410.17038",
	ARCHIVEPREFIX=	"arXiv",
	PRIMARYCLASS=	"physics.ins-det",
	REPORTNUMBER=	"CMS-TRG-19-001, CERN-EP-2024-259",
	DOI=	"10.1088/1748-0221/19/11/P11021",
	JOURNAL=	"JINST",
	VOLUME=	"19",
	PAGES=	"P11021",
	YEAR=	"2024",
}

@ARTICLE{CMS:2017yfk,
	AUTHOR=	"{CMS Collaboration}",
	TITLE=	"Particle-flow reconstruction and global event description with the {CMS} detector",
	EPRINT=	"1706.04965",
	ARCHIVEPREFIX=	"arXiv",
	PRIMARYCLASS=	"physics.ins-det",
	REPORTNUMBER=	"CMS-PRF-14-001, CERN-EP-2017-110",
	DOI=	"10.1088/1748-0221/12/10/P10003",
	JOURNAL=	"JINST",
	VOLUME=	"12",
	PAGES=	"P10003",
	YEAR=	"2017",
}

@ARTICLE{CMS:2020uim,
	AUTHOR=	"{CMS Collaboration}",
	TITLE=	"Electron and photon reconstruction and identification with the {CMS} experiment at the {CERN LHC}",
	EPRINT=	"2012.06888",
	ARCHIVEPREFIX=	"arXiv",
	PRIMARYCLASS=	"hep-ex",
	REPORTNUMBER=	"CMS-EGM-17-001, CERN-EP-2020-219",
	DOI=	"10.1088/1748-0221/16/05/P05014",
	JOURNAL=	"JINST",
	VOLUME=	"16",
	PAGES=	"P05014",
	YEAR=	"2021",
}

@ARTICLE{Sirunyan:2018fpa,
	AUTHOR=	"{CMS Collaboration}",
	TITLE=	"Performance of the {CMS} muon detector and muon reconstruction with proton-proton collisions at $\sqrt{s}=$ 13 {TeV}",
	EPRINT=	"1804.04528",
	ARCHIVEPREFIX=	"arXiv",
	PRIMARYCLASS=	"physics.ins-det",
	DOI=	"10.1088/1748-0221/13/06/P06015",
	JOURNAL=	"JINST",
	VOLUME=	"13",
	PAGES=	"P06015",
	YEAR=	"2018",
}

@ARTICLE{Khachatryan:2015dfa,
	AUTHOR=	"{CMS Collaboration}",
	TITLE=	"Reconstruction and identification of $\tau$ lepton decays to hadrons and $\nu_\tau$ at {CMS}",
	JOURNAL=	"JINST",
	VOLUME=	"11",
	YEAR=	"2016",
	PAGES=	"P01019",
	DOI=	"10.1088/1748-0221/11/01/P01019",
	EPRINT=	"1510.07488",
	ARCHIVEPREFIX=	"arXiv",
	PRIMARYCLASS=	"physics.ins-det",
	REPORTNUMBER=	"CMS-TAU-14-001, CERN-PH-EP-2015-261",
	SLACCITATION=	"%%CITATION = ARXIV:1510.07488;%%",
}

@ARTICLE{deeptau,
	AUTHOR=	"{CMS Collaboration}",
	TITLE=	"Identification of hadronic tau lepton decays using a deep neural network",
	EPRINT=	"2201.08458",
	ARCHIVEPREFIX=	"arXiv",
	PRIMARYCLASS=	"hep-ex",
	DOI=	"10.1088/1748-0221/17/07/P07023",
	JOURNAL=	"JINST",
	VOLUME=	"17",
	PAGES=	"P07023",
	YEAR=	"2022",
}

@ARTICLE{Cacciari:2008gp,
	AUTHOR=	"Cacciari, Matteo and Salam, Gavin P. and Soyez, Gregory",
	TITLE=	"The anti-\kt jet clustering algorithm",
	JOURNAL=	"JHEP",
	VOLUME=	"04",
	YEAR=	"2008",
	PAGES=	"063",
	DOI=	"10.1088/1126-6708/2008/04/063",
	EPRINT=	"0802.1189",
	ARCHIVEPREFIX=	"arXiv",
	PRIMARYCLASS=	"hep-ex",
}

@ARTICLE{Cacciari:fastjet1,
	AUTHOR=	"Matteo Cacciari and Gavin P. Salam and Gregory Soyez",
	TITLE=	"FastJet user manual",
	JOURNAL=	"Eur. Phys. J. C",
	VOLUME=	"72",
	PAGES=	"1896",
	EPRINT=	"hep-ph/1111.6097",
	ARCHIVEPREFIX=	"arXiv",
	PRIMARYCLASS=	"hep-ph",
	DOI=	"10.1140/epjc/s10052-012-1896-2",
	YEAR=	"2012",
}

@ARTICLE{Khachatryan:2016kdb,
	AUTHOR=	"{CMS Collaboration}",
	TITLE=	"Jet energy scale and resolution in the {CMS} experiment in {$\Pp\Pp$} collisions at 8 {TeV}",
	JOURNAL=	"JINST",
	VOLUME=	"12",
	YEAR=	"2017",
	PAGES=	"P02014",
	DOI=	"10.1088/1748-0221/12/02/P02014",
	EPRINT=	"1607.03663",
	ARCHIVEPREFIX=	"arXiv",
	PRIMARYCLASS=	"hep-ex",
	REPORTNUMBER=	"CMS-JME-13-004, CERN-PH-EP-2015-305",
	SLACCITATION=	"%%CITATION = ARXIV:1607.03663;%%",
}

@TECHREPORT{CMS:2017wyc,
	AUTHOR=	"{CMS Collaboration}",
	TITLE=	"Jet algorithms performance in 13 {TeV} data",
	URL=	"https://cds.cern.ch/record/2256875",
	TYPE=	"CMS Physics Analysis Summary",
	NUMBER=	"CMS-PAS-JME-16-003",
	YEAR=	"2017",
}

@ARTICLE{Sirunyan:2020foa,
	AUTHOR=	"{CMS Collaboration}",
	TITLE=	"Pileup mitigation at {CMS} in 13 {TeV} data",
	JOURNAL=	"JINST",
	VOLUME=	"15",
	PAGES=	"P09018",
	YEAR=	"2020",
	EPRINT=	"2003.00503",
	ARCHIVEPREFIX=	"arXiv",
	PRIMARYCLASS=	"hep-ex",
	REPORTNUMBER=	"CMS-JME-18-001, CERN-EP-2020-017",
	DOI=	"10.1088/1748-0221/15/09/p09018",
}

@ARTICLE{puppi,
	AUTHOR=	"Berteloni, D. and Harris P. and Low, M. and Tran, N.",
	TITLE=	"Pileup per particle identification",
	JOURNAL=	"JHEP",
	VOLUME=	"10",
	YEAR=	"2014",
	DOI=	"10.1007/JHEP10(2014)059",
	EPRINT=	"1407.6013",
	ARCHIVEPREFIX=	"arXiv",
	PRIMARYCLASS=	"hep-ph",
}

@ARTICLE{Bols:2020bkb,
	AUTHOR=	"Bols, Emil and Kieseler, Jan and Verzetti, Mauro and Stoye, Markus and Stakia, Anna",
	TITLE=	"Jet flavour classification using {DeepJet}",
	EPRINT=	"2008.10519",
	ARCHIVEPREFIX=	"arXiv",
	PRIMARYCLASS=	"hep-ex",
	DOI=	"10.1088/1748-0221/15/12/P12012",
	JOURNAL=	"JINST",
	VOLUME=	"15",
	PAGES=	"P12012",
	YEAR=	"2020",
}

@TECHREPORT{CMS-DP-2023-005,
	AUTHOR=	"{CMS Collaboration}",
	TITLE=	"Performance summary of {AK4} jet b tagging with data from proton-proton collisions at $\sqrt{s}$ = 13 {TeV} with the {CMS} detector",
	URL=	"https://cds.cern.ch/record/2854609",
	TYPE=	"CMS Detector Performance Summary",
	NUMBER=	"CMS-DP-2023-005",
	YEAR=	"2023",
}

@ARTICLE{Dasgupta:2013ihk,
	AUTHOR=	"Dasgupta, Mrinal and Fregoso, Alessandro and Marzani, Simone and Salam, Gavin P.",
	TITLE=	"Towards an understanding of jet substructure",
	JOURNAL=	"JHEP",
	VOLUME=	"09",
	YEAR=	"2013",
	PAGES=	"029",
	DOI=	"10.1007/JHEP09(2013)029",
	EPRINT=	"1307.0007",
	ARCHIVEPREFIX=	"arXiv",
	PRIMARYCLASS=	"hep-ph",
	REPORTNUMBER=	"CERN-PH-TH-2013-145, DCPT-13-86, IPPP-13-43, LPN13-036, MAN-HEP-2013-12",
	SLACCITATION=	"%%CITATION = ARXIV:1307.0007;%%",
}

@ARTICLE{Butterworth:2008iy,
	AUTHOR=	"Butterworth, Jonathan M. and Davison, Adam R. and Rubin, Mathieu and Salam, Gavin P.",
	TITLE=	"Jet substructure as a new {Higgs} search channel at the {LHC}",
	JOURNAL=	"Phys. Rev. Lett.",
	VOLUME=	"100",
	YEAR=	"2008",
	PAGES=	"242001",
	DOI=	"10.1103/PhysRevLett.100.242001",
	EPRINT=	"0802.2470",
	ARCHIVEPREFIX=	"arXiv",
	PRIMARYCLASS=	"hep-ph",
	SLACCITATION=	"%%CITATION = ARXIV:0802.2470;%%",
}

@ARTICLE{msd,
	AUTHOR=	"Larkoski, Andrew J. and Marzani, Simone and Soyez, Gregory and Thaler, Jesse",
	TITLE=	"Soft drop",
	EPRINT=	"1402.2657",
	ARCHIVEPREFIX=	"arXiv",
	PRIMARYCLASS=	"hep-ph",
	REPORTNUMBER=	"MIT-CTP-4531, DCPT-14-24, IPPP-14-12",
	DOI=	"10.1007/JHEP05(2014)146",
	JOURNAL=	"JHEP",
	VOLUME=	"05",
	PAGES=	"146",
	YEAR=	"2014",
}

@ARTICLE{Sirunyan:2020lcu,
	AUTHOR=	"{CMS Collaboration}",
	TITLE=	"Identification of heavy, energetic, hadronically decaying particles using machine-learning techniques",
	EPRINT=	"2004.08262",
	ARCHIVEPREFIX=	"arXiv",
	PRIMARYCLASS=	"hep-ex",
	REPORTNUMBER=	"CMS-JME-18-002, CERN-EP-2020-037",
	DOI=	"10.1088/1748-0221/15/06/P06005",
	JOURNAL=	"JINST",
	VOLUME=	"15",
	PAGES=	"P06005",
	YEAR=	"2020",
}

@ARTICLE{Sirunyan:2019kia,
	AUTHOR=	"{CMS Collaboration}",
	TITLE=	"Performance of missing transverse momentum reconstruction in proton-proton collisions at $\sqrt{s} = 13$\,{TeV} using the {CMS} detector",
	JOURNAL=	"JINST",
	VOLUME=	"14",
	YEAR=	"2019",
	PAGES=	"P07004",
	DOI=	"10.1088/1748-0221/14/07/P07004",
	EPRINT=	"1903.06078",
	ARCHIVEPREFIX=	"arXiv",
	PRIMARYCLASS=	"hep-ex",
	REPORTNUMBER=	"CMS-JME-17-001, CERN-EP-2018-335",
	SLACCITATION=	"%%CITATION = ARXIV:1903.06078;%%",
}

@ARTICLE{Sjostrand:2014zea,
	AUTHOR=	"Sj{\"o}strand, Torbj{\"o}rn and Ask, Stefan and Christiansen, Jesper R. and Corke, Richard and Desai, Nishita and Ilten, Philip and Mrenna, Stephen and Prestel, Stefan and Rasmussen, Christine O. and Skands, Peter Z.",
	TITLE=	"An Introduction to {PYTHIA} 8.2",
	JOURNAL=	"Comput. Phys. Commun.",
	VOLUME=	"191",
	YEAR=	"2015",
	PAGES=	"159",
	DOI=	"10.1016/j.cpc.2015.01.024",
	EPRINT=	"1410.3012",
	ARCHIVEPREFIX=	"arXiv",
	PRIMARYCLASS=	"hep-ph",
	REPORTNUMBER=	"LU-TP-14-36, MCNET-14-22, CERN-PH-TH-2014-190, FERMILAB-PUB-14-316-CD, DESY-14-178, SLAC-PUB-16122",
	SLACCITATION=	"%%CITATION = ARXIV:1410.3012;%%",
}

@ARTICLE{CMS:2019csb,
	AUTHOR=	"{CMS Collaboration}",
	TITLE=	"Extraction and validation of a new set of {CMS PYTHIA8} tunes from underlying-event measurements",
	EPRINT=	"1903.12179",
	ARCHIVEPREFIX=	"arXiv",
	PRIMARYCLASS=	"hep-ex",
	REPORTNUMBER=	"CMS-GEN-17-001, CERN-EP-2019-007",
	DOI=	"10.1140/epjc/s10052-019-7499-4",
	JOURNAL=	"Eur. Phys. J. C",
	VOLUME=	"80",
	PAGES=	"4",
	YEAR=	"2020",
}

@ARTICLE{Agostinelli:2002hh,
	AUTHOR=	"Agostinelli, S. and others",
	COLLABORATION=	"GEANT4",
	TITLE=	"{\GEANTfour}---a simulation toolkit",
	JOURNAL=	"Nucl. Instrum. Meth. A",
	VOLUME=	"506",
	YEAR=	"2003",
	PAGES=	"250",
	DOI=	"10.1016/S0168-9002(03)01368-8",
	SLACCITATION=	"%%CITATION = NUIMA,A506,250;%%",
}

@ARTICLE{Ball2015,
	AUTHOR=	"Ball, Richard D. and others",
	COLLABORATION=	"NNPDF",
	TITLE=	"Parton distributions for the {LHC Run II}",
	EPRINT=	"1410.8849",
	ARCHIVEPREFIX=	"arXiv",
	PRIMARYCLASS=	"hep-ph",
	REPORTNUMBER=	"EDINBURGH-2014-15, IFUM-1034-FT, CERN-PH-TH-2013-253, OUTP-14-11P, CAVENDISH-HEP-14-11",
	DOI=	"10.1007/JHEP04(2015)040",
	JOURNAL=	"JHEP",
	VOLUME=	"04",
	PAGES=	"040",
	YEAR=	"2015",
}

@ARTICLE{Ball:2017nwa,
	AUTHOR=	"Ball, Richard D. and others",
	TITLE=	"Parton distributions from high-precision collider data",
	COLLABORATION=	"NNPDF",
	JOURNAL=	"Eur. Phys. J.",
	VOLUME=	"77",
	YEAR=	"2017",
	PAGES=	"663",
	DOI=	"10.1140/epjc/s10052-017-5199-5",
	EPRINT=	"1706.00428",
	ARCHIVEPREFIX=	"arXiv",
	PRIMARYCLASS=	"hep-ph",
	REPORTNUMBER=	"CAVENDISH-HEP-17-06, CERN-TH-2017-077, EDINBURGH-2017-08, NIKHEF-2017-006, OUTP-17-04P, TIF-UNIMI-2017-3",
	SLACCITATION=	"%%CITATION = ARXIV:1706.00428;%%",
}

@ARTICLE{MADGRAPH,
	AUTHOR=	"Alwall, J. and Frederix, R. and Frixione, S. and Hirschi, V. and Maltoni, F. and Mattelaer, O. and Shao, H. -S. and Stelzer, T. and Torrielli, P. and Zaro, M.",
	TITLE=	"The automated computation of tree-level and next-to-leading order differential cross sections, and their matching to parton shower simulations",
	JOURNAL=	"JHEP",
	VOLUME=	"07",
	YEAR=	"2014",
	PAGES=	"079",
	DOI=	"10.1007/JHEP07(2014)079",
	EPRINT=	"1405.0301",
	ARCHIVEPREFIX=	"arXiv",
	PRIMARYCLASS=	"hep-ph",
	REPORTNUMBER=	"CERN-PH-TH-2014-064, CP3-14-18, LPN14-066, MCNET-14-09, ZU-TH-14-14",
	SLACCITATION=	"%%CITATION = ARXIV:1405.0301;%%",
}

@ARTICLE{Mangano:2006rw,
	AUTHOR=	"Mangano, Michelangelo L. and Moretti, Mauro and Piccinini, Fulvio and Treccani, Michele",
	TITLE=	"Matching matrix elements and shower evolution for top-quark production in hadronic collisions",
	JOURNAL=	"JHEP",
	VOLUME=	"01",
	YEAR=	"2007",
	PAGES=	"013",
	DOI=	"10.1088/1126-6708/2007/01/013",
	EPRINT=	"hep-ph/0611129",
	ARCHIVEPREFIX=	"arXiv",
	PRIMARYCLASS=	"hep-ph",
	REPORTNUMBER=	"CERN-PH-TH-2006-232, FNT-T-2006-09",
	SLACCITATION=	"%%CITATION = HEP-PH/0611129;%%",
}

@ARTICLE{Frederix:2012ps,
	AUTHOR=	"Frederix, Rikkert and Frixione, Stefano",
	TITLE=	"Merging meets matching in {MC@NLO}",
	JOURNAL=	"JHEP",
	VOLUME=	"12",
	YEAR=	"2012",
	PAGES=	"061",
	DOI=	"10.1007/JHEP12(2012)061",
	EPRINT=	"1209.6215",
	ARCHIVEPREFIX=	"arXiv",
	PRIMARYCLASS=	"hep-ph",
	REPORTNUMBER=	"CERN-PH-TH-2012-247, ZU-TH-21-12",
	SLACCITATION=	"%%CITATION = ARXIV:1209.6215;%%",
}

@ARTICLE{Czakon:2013goa,
	AUTHOR=	"Czakon, Micha{\l} and Fiedler, Paul and Mitov, Alexander",
	TITLE=	"Total Top-Quark Pair-Production Cross Section at Hadron Colliders Through {$O(\alpha^4_S)$}",
	EPRINT=	"1303.6254",
	ARCHIVEPREFIX=	"arXiv",
	PRIMARYCLASS=	"hep-ph",
	REPORTNUMBER=	"CERN-PH-TH-2013-056, TTK-13-08",
	DOI=	"10.1103/PhysRevLett.110.252004",
	JOURNAL=	"Phys. Rev. Lett.",
	VOLUME=	"110",
	PAGES=	"252004",
	YEAR=	"2013",
}

@ARTICLE{Nason:2004rx,
	AUTHOR=	"Nason, Paolo",
	TITLE=	"A new method for combining {NLO QCD} with shower {Monte Carlo} algorithms",
	EPRINT=	"hep-ph/0409146",
	ARCHIVEPREFIX=	"arXiv",
	REPORTNUMBER=	"BICOCCA-FT-04-11",
	DOI=	"10.1088/1126-6708/2004/11/040",
	JOURNAL=	"JHEP",
	VOLUME=	"11",
	PAGES=	"040",
	YEAR=	"2004",
}

@ARTICLE{Frixione:2007vw,
	AUTHOR=	"Frixione, Stefano and Nason, Paolo and Oleari, Carlo",
	TITLE=	"Matching {NLO QCD} computations with Parton Shower simulations: the {POWHEG} method",
	EPRINT=	"0709.2092",
	ARCHIVEPREFIX=	"arXiv",
	PRIMARYCLASS=	"hep-ph",
	REPORTNUMBER=	"BICOCCA-FT-07-9, GEF-TH-21-2007",
	DOI=	"10.1088/1126-6708/2007/11/070",
	JOURNAL=	"JHEP",
	VOLUME=	"11",
	PAGES=	"070",
	YEAR=	"2007",
}

@ARTICLE{Alioli:2010xd,
	AUTHOR=	"Alioli, Simone and Nason, Paolo and Oleari, Carlo and Re, Emanuele",
	TITLE=	"A general framework for implementing {NLO} calculations in shower {Monte Carlo} programs: the {POWHEG BOX}",
	EPRINT=	"1002.2581",
	ARCHIVEPREFIX=	"arXiv",
	PRIMARYCLASS=	"hep-ph",
	REPORTNUMBER=	"DESY-10-018, SFB-CPP-10-22, IPPP-10-11, DCPT-10-22",
	DOI=	"10.1007/JHEP06(2010)043",
	JOURNAL=	"JHEP",
	VOLUME=	"06",
	PAGES=	"043",
	YEAR=	"2010",
}

@ARTICLE{Alioli:2009je,
	AUTHOR=	"Alioli, Simone and Nason, Paolo and Oleari, Carlo and Re, Emanuele",
	TITLE=	"{NLO} single-top production matched with shower in {POWHEG}: $s$- and $t$-channel contributions",
	JOURNAL=	"JHEP",
	VOLUME=	"09",
	YEAR=	"2009",
	PAGES=	"111",
	DOI=	"10.1088/1126-6708/2009/09/111",
	NOTE=	"[Erratum: \doi{10.1007/JHEP02(2010)011}]",
	EPRINT=	"0907.4076",
	ARCHIVEPREFIX=	"arXiv",
	PRIMARYCLASS=	"hep-ph",
	SLACCITATION=	"%%CITATION = ARXIV:0907.4076;%%",
}

@ARTICLE{Re:2010bp,
	AUTHOR=	"Re, Emanuele",
	TITLE=	"Single-top {Wt}-channel production matched with parton showers using the {POWHEG} method",
	JOURNAL=	"Eur. Phys. J. C",
	VOLUME=	"71",
	YEAR=	"2011",
	PAGES=	"1547",
	DOI=	"10.1140/epjc/s10052-011-1547-z",
	EPRINT=	"1009.2450",
	ARCHIVEPREFIX=	"arXiv",
	PRIMARYCLASS=	"hep-ph",
	REPORTNUMBER=	"IPPP-10-74, DCPT-10-148",
	SLACCITATION=	"%%CITATION = ARXIV:1009.2450;%%",
}

@ARTICLE{Aliev:2010zk,
	AUTHOR=	"Aliev, M. and Lacker, H. and Langenfeld, U. and Moch, S. and Uwer, P. and Wiedermann, M.",
	ARCHIVEPREFIX=	"arXiv",
	DOI=	"10.1016/j.cpc.2010.12.040",
	EPRINT=	"1007.1327",
	JOURNAL=	"Comput. Phys. Commun.",
	PAGES=	"1034",
	PRIMARYCLASS=	"hep-ph",
	REPORTNUMBER=	"DESY-10-091, HU-EP-10-33, SFB-CPP-10-60",
	TITLE=	"{HATHOR}: {HAdronic} {Top} and {Heavy} quarks cr{O}ss section calculato{R}",
	VOLUME=	"182",
	YEAR=	"2011",
}

@ARTICLE{Kant:2014oha,
	AUTHOR=	"Kant, P. and Kind, O. M. and Kintscher, T. and Lohse, T. and Martini, T. and M{\"}olbitz, S. and Rieck, P. and Uwer, P.",
	ARCHIVEPREFIX=	"arXiv",
	DOI=	"10.1016/j.cpc.2015.02.001",
	EPRINT=	"1406.4403",
	JOURNAL=	"Comput. Phys. Commun.",
	PAGES=	"74",
	PRIMARYCLASS=	"hep-ph",
	REPORTNUMBER=	"HU-EP-14-22",
	TITLE=	"{HATHOR} for single top-quark production: Updated predictions and uncertainty estimates for single top-quark production in hadronic collisions",
	VOLUME=	"191",
	YEAR=	"2015",
}

@ARTICLE{Gehrmann:2014fva,
	AUTHOR=	"Gehrmann, T. and Grazzini, M. and Kallweit, S. and Maierh{\"o}fer, P. and von Manteuffel, A. and Pozzorini, S. and Rathlev, D. and Tancredi, L.",
	TITLE=	"{$\PW^+\PW^-$} Production at Hadron Colliders in Next to Next to Leading Order {QCD}",
	JOURNAL=	"Phys. Rev. Lett.",
	VOLUME=	"113",
	YEAR=	"2014",
	PAGES=	"212001",
	DOI=	"10.1103/PhysRevLett.113.212001",
	EPRINT=	"1408.5243",
	ARCHIVEPREFIX=	"arXiv",
	PRIMARYCLASS=	"hep-ph",
	REPORTNUMBER=	"ZU-TH-29-14, MITP-14-053",
	SLACCITATION=	"%%CITATION = ARXIV:1408.5243;%%",
}

@ARTICLE{Campbell:1999ah,
	AUTHOR=	"Campbell, John M. and Ellis, R. Keith",
	TITLE=	"An Update on vector boson pair production at hadron colliders",
	JOURNAL=	"Phys. Rev. D",
	VOLUME=	"60",
	YEAR=	"1999",
	PAGES=	"113006",
	DOI=	"10.1103/PhysRevD.60.113006",
	EPRINT=	"hep-ph/9905386",
	ARCHIVEPREFIX=	"arXiv",
	PRIMARYCLASS=	"hep-ph",
	REPORTNUMBER=	"FERMILAB-PUB-99-146-T",
	SLACCITATION=	"%%CITATION = HEP-PH/9905386;%%",
}

@ARTICLE{DMTheory,
	AUTHOR=	"Lindert, J. M. and others",
	TITLE=	"Precise predictions for {V}+jets dark matter backgrounds",
	JOURNAL=	"Eur. Phys. J.",
	VOLUME=	"77",
	YEAR=	"2017",
	PAGES=	"829",
	DOI=	"10.1140/epjc/s10052-017-5389-1",
	EPRINT=	"1705.04664",
	ARCHIVEPREFIX=	"arXiv",
	PRIMARYCLASS=	"hep-ph",
	REPORTNUMBER=	"CERN-TH-2017-102, CERN-LPCC-2017-02, IPPP-17-38, FERMILAB-PUB-17-152-T, ZU--TH-12-17",
	SLACCITATION=	"%%CITATION = ARXIV:1705.04664;%%",
}

@ARTICLE{CMS:2021xjt,
	AUTHOR=	"{CMS Collaboration}",
	TITLE=	"Precision luminosity measurement in proton-proton collisions at $\sqrt{s} =$ 13 {TeV} in 2015 and 2016 at {CMS}",
	EPRINT=	"2104.01927",
	ARCHIVEPREFIX=	"arXiv",
	PRIMARYCLASS=	"hep-ex",
	REPORTNUMBER=	"CMS-LUM-17-003, CERN-EP-2021-033",
	DOI=	"10.1140/epjc/s10052-021-09538-2",
	JOURNAL=	"Eur. Phys. J. C",
	VOLUME=	"81",
	PAGES=	"800",
	YEAR=	"2021",
}

@TECHREPORT{CMS-PAS-LUM-17-004,
	TITLE=	"{CMS} luminosity measurement for the 2017 data-taking period at $\sqrt{s}$ = 13 {TeV}",
	AUTHOR=	"{CMS Collaboration}",
	URL=	"https://cds.cern.ch/record/2621960/",
	TYPE=	"CMS Physics Analysis Summary",
	NUMBER=	"CMS-PAS-LUM-17-004",
	YEAR=	"2018",
}

@TECHREPORT{CMS-PAS-LUM-18-002,
	TITLE=	"{CMS} luminosity measurement for the 2018 data-taking period at $\sqrt{s}$ = 13 {TeV}",
	AUTHOR=	"{CMS Collaboration}",
	URL=	"https://cds.cern.ch/record/2676164/",
	TYPE=	"CMS Physics Analysis Summary",
	NUMBER=	"CMS-PAS-LUM-18-002",
	YEAR=	"2019",
}

@ARTICLE{Khachatryan:2014gga,
	AUTHOR=	"{CMS Collaboration}",
	TITLE=	"Performance of the {CMS} missing transverse momentum reconstruction in {$\Pp\Pp$} data at $\sqrt{s}$ = 8 {TeV}",
	JOURNAL=	"JINST",
	VOLUME=	"10",
	YEAR=	"2015",
	PAGES=	"P02006",
	DOI=	"10.1088/1748-0221/10/02/P02006",
	EPRINT=	"1411.0511",
	ARCHIVEPREFIX=	"arXiv",
	PRIMARYCLASS=	"physics.ins-det",
	REPORTNUMBER=	"CMS-JME-13-003, CERN-PH-EP-2014-246",
	SLACCITATION=	"%%CITATION = ARXIV:1411.0511;%%",
}

@ARTICLE{Sirunyan:2017jix,
	AUTHOR=	"{CMS Collaboration}",
	TITLE=	"Search for new physics in final states with an energetic jet or a hadronically decaying {$\PW$} or {$\PZ$} boson and transverse momentum imbalance at {$\sqrt{s}=13\TeV$}",
	JOURNAL=	"Phys. Rev. D",
	VOLUME=	"97",
	YEAR=	"2018",
	PAGES=	"092005",
	DOI=	"10.1103/PhysRevD.97.092005",
	EPRINT=	"1712.02345",
	ARCHIVEPREFIX=	"arXiv",
	PRIMARYCLASS=	"hep-ex",
	REPORTNUMBER=	"CMS-EXO-16-048, CERN-EP-2017-294",
	SLACCITATION=	"%%CITATION = ARXIV:1712.02345;%%",
}

@INPROCEEDINGS{Conway:2011in,
	AUTHOR=	"Conway, J. S.",
	TITLE=	"Incorporating Nuisance Parameters in Likelihoods for Multisource Spectra",
	BOOKTITLE=	"{PHYSTAT 2011}",
	EPRINT=	"1103.0354",
	ARCHIVEPREFIX=	"arXiv",
	PRIMARYCLASS=	"physics.data-an",
	DOI=	"10.5170/CERN-2011-006.115",
	PAGES=	"115",
	YEAR=	"2011",
}

@ARTICLE{CLS1,
	AUTHOR=	"Junk, Thomas",
	TITLE=	"Confidence level computation for combining searches with small statistics",
	JOURNAL=	"Nucl. Instrum. Meth. A",
	VOLUME=	"434",
	YEAR=	"1999",
	PAGES=	"435",
	DOI=	"10.1016/S0168-9002(99)00498-2",
	EPRINT=	"hep-ex/9902006",
	ARCHIVEPREFIX=	"arXiv",
	PRIMARYCLASS=	"hep-ex",
	REPORTNUMBER=	"CARLETON-OPAL-PHYS-99-01, CERN-EP-99-041",
	SLACCITATION=	"%%CITATION = HEP-EX/9902006;%%",
}

@ARTICLE{CLS2,
	AUTHOR=	"Read, Alexander L.",
	TITLE=	"Presentation of search results: The {CL$_{\text{s}}$} technique",
	JOURNAL=	"J. Phys. G",
	VOLUME=	"28",
	YEAR=	"2002",
	PAGES=	"2693",
	DOI=	"10.1088/0954-3899/28/10/313",
	SLACCITATION=	"%%CITATION = JPAGA,G28,2693;%%",
}

@ARTICLE{Cowan:2010js,
	AUTHOR=	"Cowan, Glen and Cranmer, Kyle and Gross, Eilam and Vitells, Ofer",
	TITLE=	"Asymptotic formulae for likelihood-based tests of new physics",
	JOURNAL=	"Eur. Phys. J. C",
	VOLUME=	"71",
	YEAR=	"2011",
	PAGES=	"1554",
	DOI=	"10.1140/epjc/s10052-011-1554-0",
	NOTE=	"[Erratum: \DOI{10.1140/epjc/s10052-013-2501-z}]",
	EPRINT=	"1007.1727",
	ARCHIVEPREFIX=	"arXiv",
	PRIMARYCLASS=	"physics.data-an",
	SLACCITATION=	"%%CITATION = ARXIV:1007.1727;%%",
}
